\documentclass[12pt]{article}
\usepackage{epsfig,amsfonts,amssymb}
\usepackage{hyperref}
\usepackage{cite}
\input epsf.sty
\usepackage{cite}

\def\ZZZ{{\hbox{ Z\kern-1.6mm Z}}}
\def\RRR{{\hbox{ R\kern-2.4mm R}}}
\def\CCC{{\hbox{ C\kern-2.0mm C}}}
\def\zzz{{\hbox{z\kern-1mm z}}}

\newcommand{\vt}{\vartheta}

\newcommand{\qeq}{{\hbox{=\kern-2.3mm ? \kern.5mm }}}
\renewcommand{\qeq}{=}

\newcommand{\BB}{{\cal B}}
\newcommand{\BBB}{{\cal B}}

\newcommand{\AAA}{{\cal A}}

\newcommand{\CC}{{\cal C}}

\newcommand{\OO}{{\cal O}}

\newcommand{\wt}{\widetilde}
\newcommand{\wh}{\widehat}

\newcommand{\NN}{{\cal N}}

\newcommand{\ppp}{\prime\prime}

\newcommand{\cp}{\check\Phi}
\newcommand{\crh}{\check\rho}
\newcommand{\cs}{\check\sigma}
\newcommand{\cv}{\check v}
\newcommand{\com}{\check\Omega}

\newcommand{\be}{\begin{equation}}
\newcommand{\ee}{\end{equation}}
\newcommand{\ben}{\begin{eqnarray}\displaystyle}
\newcommand{\een}{\end{eqnarray}}

\newcommand{\bea}[1]{\begin{eqnarray}\label{#1} }
\newcommand{\eea}{\end{eqnarray}}

\newcommand{\refb}[1]{(\ref{#1})}

\newcommand{\sectiono}[1]{\section{#1}\setcounter{equation}{0}}
\newcommand{\subsectiono}[1]{\subsection{#1}\setcounter{equation}{0}}

\def\one{{\hbox{ 1\kern-.8mm l}}}
\def\zero{{\hbox{ 0\kern-1.5mm 0}}}

\renewcommand{\theequation}{\thesection.\arabic{equation}}

\begin{document}

\baselineskip 24pt

\begin{center}
{\Large \bf
Generalities of Quarter BPS Dyon Partition Function and
Dyons of Torsion Two}

\end{center}

\vskip .6cm
\medskip

\vspace*{4.0ex}

\baselineskip=18pt

\centerline{\large \rm  Shamik Banerjee,  Ashoke Sen
and Yogesh K. Srivastava}

\vspace*{4.0ex}

\centerline{\large \it Harish-Chandra Research Institute}

\centerline{\large \it  Chhatnag Road, Jhusi,
Allahabad 211019, INDIA}

\vspace*{1.0ex}
\centerline{E-mail:  bshamik, sen, yogesh@mri.ernet.in}

\vspace*{5.0ex}

\centerline{\bf Abstract} \bigskip

We propose a general set of constraints on the partition function
of quarter BPS dyons in any $\NN=4$ supersymmetric string theory
by drawing insight from known examples, and study the consequences
of this proposal. The  main ingredients of our analysis are
duality symmetries, wall crossing formula and black hole 
entropy. We use our analysis to
constrain the dyon partition function for two hitherto unknown
cases -- the partition function of dyons of torsion 
two (\i.e. gcd($Q\wedge P$)=2) in heterotic
string theory on $T^6$ and the partition function of dyons carrying
untwisted sector electric charge in $\ZZZ_2$ CHL model. 
With the help of these constraints we propose a candidate for
the partition function of dyons of torsion two in
heterotic string theory on $T^6$. This leads to a novel wall
crossing formula for decay of quarter BPS dyons into half BPS
dyons with non-primitive charge vectors. In an appropriate
limit the proposed formula reproduces the known result
for the spectrum of torsion two dyons in gauge theory.

\vfill \eject

\baselineskip=18pt

\tableofcontents

\sectiono{Introduction and summary} \label{s1}

The partition function of quarter BPS dyons is now known in
a variety of $\NN=4$ supersymmetric string
theories\cite{9607026,0412287,0505094,
0506249,0508174,0510147,0602254,
0603066,0605210,0607155,0609109,0612011,0702141,
0702150,0705.1433,0705.3874,0706.2363,0708.1270}. 
Generalization of these results to a class of $\NN=2$ supersymmetric
string theories have also been proposed\cite{0711.1971}.
Our goal in this paper is to draw
insight from these known results to postulate the general structure
of dyon partition function for any class of quarter BPS dyons
in any $\NN=4$ supersymmetric string
theory.

The results of our analysis can be summarized as 
follows.
\begin{enumerate}
\item {\bf Definition of the partition function:}
Let $(Q,P)$ denote the electric and magnetic charges carried
by a dyon, and $Q^2$, $P^2$ and $Q\cdot P$ be the T-duality
invariant quadratic forms constructed from these 
charges.\footnote{Irrespective of what description we are using,
we shall denote by S-duality transformation the symmetry that acts
on the complex scalar belonging to the gravity multiplet. In heterotic
string compactification this would correspond to the axion-dilaton
modulus. On the other hand T-duality will denote the symmetry that
acts on the matter multiplet scalars. In the heterotic description
these scalars arise from the components of the metric, anti-symmetric
tensor fields and gauge fields along the compact directions.}
In order to define the dyon partition function we first need
to identify a suitable infinite
subset $\BBB$ of dyons in the theory with the
property that if we have two pairs of charges $(Q,P)\in\BBB$ 
and $(Q',P')\in\BBB$
with $Q^2=Q^{\prime2}$, $P^2=P^{\prime2}$ and $Q\cdot P
=Q'\cdot P'$, then they must be related by a
T-duality transformation. Furthermore given a pair of charge vectors
$(Q,P)\in\BBB$, all other pairs of
charge vectors related to it by
T-duality should be elements of the set $\BBB$.
We shall generate such a set $\BBB$ by beginning with a family
$\AAA$ of charge vectors $(Q,P)$ labelled by three integers
such that $Q^2$, $P^2$ and $Q\cdot P$ are independent linear
functions of these three integers, and then define $\BBB$ to be
the set of all $(Q,P)$ which are in the T-duality orbit of 
the set $\AAA$.
We denote by $d(Q,P)$ the degeneracy, -- or more
precisely an index measuring the number of bosonic supermultiplets
minus  the
number of fermionic supermultiplets -- of quarter BPS
dyons of charge $(Q,P)$.
Since $d(Q,P)$
should be invariant under
a T-duality transformation,  for $(Q,P)\in\BBB$ 
it should depend on $(Q,P)$ only
via the T-duality invariant combinations $Q^2$, $P^2$ and 
$Q\cdot P$:
\be \label{ei1}
d(Q,P) = f(Q^2, P^2, Q\cdot P)\, .
\ee
Note that for \refb{ei1} to hold it is necessary to choose $\BBB$ in
the way we have described. In particular if $\BBB$ 
had contained two elements with same
$Q^2$, $P^2$ and $Q\cdot P$ but not related by a T-duality 
transformation, then $d(Q,P)$ can be different for these two
elements and \refb{ei1} will not hold. Even when
$\BBB$ is chosen according to the prescription given
above, eq.\refb{ei1} cannot be strictly correct since $d(Q,P)$
could depend on the asymptotic moduli besides the charges and one
can construct more general T-duality invariants using these
moduli and the charges. Indeed, 
even though the index is not expected to change under
a continuous change in the moduli, it could jump across the walls
of marginal stability giving $d(Q,P)$ a dependence on the moduli.
The reason that we can still write eq.\refb{ei1}
is that 
it is possible to label the
different domains bounded by the walls of marginal stability 
by a set of discrete parameters $\vec c$ such 
that T-duality transformation does not change the 
parameters $\vec c$\cite{0702141,0708.1270,0801.0149}.
Physically, if a domain is bounded by $n$ walls of marginal
stability, with the $i$th wall associated with the decay
$(Q,P)\to (\alpha_i Q+\beta_i P, \gamma_i Q+\delta_i P)
+ ((1-\alpha_i) Q-\beta_i P, -\gamma_i Q+(1-\delta_i) P)$,
then $\vec c$ is the collection of the numbers $\{(\alpha_i,
\beta_i,\gamma_i,\delta_i); \, 1\le i\le n\}$. 
Due to T-duality invariance of
$\vec c$,
$d(Q,P)$ inside a given domain labelled by
$\vec c$ will be invariant under a T-duality transformation on
the charges only and will have 
the form \refb{ei1}. For different $\vec c$ the function $f$ 
will be different, \i.e.\ $f$ has a hidden $\vec c$
dependence. We now define the dyon partition function associated
with the set $\BBB$ to be\footnote{For $\NN=4$
supersymmetric $\ZZZ_N$
orbifolds reviewed in \cite{0708.1270} 
the function $\cp$ is related to
the function $\wt\Phi$ of \cite{0708.1270} by the relation
$\cp(\crh,\cs,\cv) = \wt\Phi(\wt\rho,\wt\sigma,\wt v)$ with
$(\wt\rho,\wt\sigma,\wt v)=(\cs/N, N\crh, \cv)$.}
\be\label{ei2}
{1\over \cp(\crh,\cs,\cv)} 
\equiv \sum_{Q^2, P^2, Q\cdot P}\, 
(-1)^{Q\cdot P+1}\, f(Q^2,P^2, Q\cdot P) 
e^{i\pi (\cs Q^2 + \crh P^2 
+ 2 \cv Q\cdot P)}\, ,
\ee
where the sum runs over all the distinct triplets 
$(Q^2,P^2,Q\cdot P)$ which are present in the set $\BBB$. This
relation can be inverted as
\be \label{ei3}
f(Q^2,P^2, Q\cdot P)
\propto (-1)^{Q\cdot P+1} \int_\CC
 d\crh
d\cs  d\cv \,
e^{-i\pi (\cs Q^2 + \crh P^2 
+ 2 \cv Q\cdot P)} \, 
{1\over \cp(\crh, \cs, \cv)}\, ,
\ee
where $\CC$ denotes an appropriate three dimensional subspace
of the complex $(\crh,\cs,\cv)$ space. Along the `contour'
$\CC$ the imaginary parts of $\crh$, $\cs$ and $\cv$ are fixed at
values where the sum in \refb{ei2}
converges, and the real parts of $\crh$, $\cs$ and $\cv$ vary over
an appropriate unit cell determined by the quantization laws of
$Q^2$, $P^2$ and $Q\cdot P$ inside the set $\BBB$.

In all known cases the function $f$ in different domains $\vec c$
is given by \refb{ei3} with identical integrand, but the
integration contour $\CC$ depends on the choice of $\vec c$.
Put another way, the same function $1/\cp(\crh,\cs,\cv)$ admits
different Fourier expansion in different regions in the complex
$(\crh,\cs,\cv)$ space, since a Fourier expansion that is
convergent in one region may not be convergent in another
region. The coefficients of expansion in these different regions in
the complex $(\crh,\cs,\cv)$ plane may then be regarded as
the index $f(Q^2,P^2, Q\cdot P)$ in different domains
in the asymptotic moduli space labelled by $\vec c$.
We shall assume that {\it this result holds for all sets of dyons
in all $\NN=4$ string theories.}

\item {\bf Consequences of S-duality symmetry:}
We now consider the effect of an S-duality transformation
on the set $\BBB$. A generic S-duality transformation will take 
an element of $\BBB$ to outside $\BBB$, -- we denote by
$H$ the subgroup of the S-duality group that leaves $\BBB$ 
invariant. This is the subgroup relevant for constraining the dyon
partition function associated with the set $\BBB$. Since a
generic element of $H$ takes us from one domain bounded by walls
of marginal stability to another such domain,
it relates the function $f$ for one choice of $\vec c$ to the function
$f$ for another choice of $\vec c$.
However since we have assumed that
the dyon partition function $1/\cp$ is independent
of the domain label $\vec c$, we can use invariance under $H$
to constrain the form of $\cp$.  In particular one finds that
an S-duality symmetry of the form $(Q,P)\to (aQ+bP,cQ+dP)$
with $\pmatrix{a & b\cr c & d}\in H$ gives the following constraint
on $\cp$:
\be \label{ei4}
\cp(\crh,\cs,\cv) = \cp(d^2 \crh+ b^2 \cs + 2 bd \cv,
c^2 \crh + a^2 \cs + 2 ac \cv,cd \crh+ ab \cs + (ad+bc) \cv)\, .
\ee
Defining 
\be \label{ei5}
\com=\pmatrix{\crh & \cv\cr \cv & \cs} \, ,
\ee
we can express \refb{ei4} as
\be \label{ei6}
\cp((A\com+B)(C\com +D)^{-1})
= (\det(C\com +D))^k \, \cp(\Omega)\, ,
\ee
where
\be \label{ei7}
\pmatrix{A & B\cr C & D}
=  \pmatrix{d & b &0&0 \cr c & a &0&0\cr
0&0& a & -c\cr 0&0& -b & d}\, ,
\ee
and $k$ is as yet undermined since $\det(C\Omega+D)=1$.

Besides this symmetry, quantization of $Q^2$, $P^2$
and $Q\cdot P$ within the set $\BBB$
also gives rise to some translational
symmetries of $\cp$ of the form 
$\cp(\crh,\cs,\cv)=\cp(\crh+a_1,\cs+a_2,\cv+a_3)$
with $a_1$, $a_2$, $a_3$ taking values in an appropriate set.
These can also be expressed as \refb{ei6} with
\be \label{ei8}
\pmatrix{A & B\cr C & D} = \pmatrix{1 & 0 & a_1 & a_3\cr
0 & 1 & a_3 & a_2\cr 0 & 0 & 1 & 0\cr 0 & 0 & 0 & 1}\, .
\ee

\item {\bf Wall crossing formula:} Given that the indices in
different domains in the moduli space are given by different choices of
the 3-dimensional 
integration contour in the $(\crh,\cs,\cv)$ space, 
the jump in the index as we cross a
wall of marginal stability
must be given by the residue of the integrand at
the pole(s) encountered while deforming one contour to another.
The walls across which the index jumps are the ones associated
with decays 
into a pair of half-BPS states.\footnote{For 
a certain class of dyons kinematics allows decay into a
pair of quarter BPS states or a half BPS and a quarter BPS states
on a codimension 1 subspace 
of the moduli space. These correspond to decays of the form
$(Q,P)\to (\alpha Q+\beta P, \gamma Q+\delta P) +
((1-\alpha) Q-\beta P, -\gamma Q+(1-\delta)P)$ with some of
the $\alpha,\beta,\gamma,\delta$ fractional so that we can have
$0<(\alpha\delta -\beta\gamma)<1$ and $0\le ((1-\alpha)(1-\delta)
-\beta\gamma <1$\cite{0707.3035,0707.1563}.
However a naive counting
of the number of fermion zero modes on a half BPS - quarter BPS
and quarter BPS - quarter BPS
combination suggests that there are additional fermion zero modes
besides the ones associated with the broken supersymmetry
generators. This
makes the index associated with
such a configuration vanish. 
Although a rigorous analysis of this system is
lacking at present, we shall proceed with the assumption that the
result is valid so that
such decays do not change the index. 
Otherwise the dyon partition function will have additional poles
associated with the jump in the index across these additional walls
of marginal stability.
We wish to thank F.~Denef for a discussion 
on this point. \label{foot1}} 
We can label the decay products as\cite{0702141}
\be\label{eff1}
(Q,P) \to (a_0d_0 Q - a_0b_0 P, c_0d_0 Q - c_0b_0 P) +
(-b_0c_0 Q + a_0b_0 P, - c_0d_0 Q + a_0d_0 P)\, , \ee
where $a_0$, $b_0$, $c_0$ and $d_0$ are normalized so that $a_0 d_0
- b_0 c_0=1$. In a generic situation 
$a_0$, $b_0$, $c_0$ and $d_0$ are not necessarily integers but are
constrained by the fact that 
the final charges satisfy the charge quantization laws.
In all known examples there is a specific
correlation between a wall corresponding to a given decay and the
location of the pole of the integrand that the contour crosses as we
cross the wall in the moduli space. The location of the pole
associated with the decay \refb{eff1} is given by:
\be \label{eff2}
\crh c_0d_0 + \cs a_0b_0 + \cv (a_0d_0 + b_0c_0) = 0\, .
\ee 
We shall
assume that {\it this formula continues to hold in all cases.} This
then relates the jump in the index across a given wall of
marginal stability
to the residue of the partition function at a specific pole. An
explicit
choice of moduli dependent 
contour that satisfies this requirement can be
found by generalizing the result of Cheng
and Verlinde\cite{0706.2363} to generic quarter BPS
dyons in generic $\NN=4$ supersymmetric
string theories:
\ben \label{echoiceint}
\Im(\crh) &=& \Lambda \, \left({|\tau|^2\over \tau_2} +
{Q_R^2 \over \sqrt{Q_R^2 P_R^2 - (Q_R\cdot P_R)^2}}\right)\, ,
\nonumber \\
\Im(\cs) &=& \Lambda \, \left({1\over \tau_2} +
{P_R^2 \over \sqrt{Q_R^2 P_R^2 - (Q_R\cdot P_R)^2}}\right)\, ,
\nonumber \\
\Im(\cv) &=& -\Lambda \, \left({\tau_1\over \tau_2} +
{Q_R\cdot P_R \over \sqrt{Q_R^2 P_R^2 - (Q_R\cdot P_R)^2}}
\right)\, ,
\een
where $\Lambda$ is a large positive number, $\Im(z)$ denotes
the imaginary part of $z$,
\be \label{edefcrint}
Q_R^2 = Q^T (M+L)Q, \quad P_R^2=P^T (M+L)P, \quad
Q_R\cdot P_R = Q^T(M+L)P\, ,
\ee
$\tau\equiv \tau_1+i\tau_2$ denotes the asymptotic value
of the axion-dilaton moduli which
belong to the gravity multiplet and $M$ is the 
asymptotic value of the symmetric matrix
valued moduli field of the matter multiplet 
satisfying $MLM^T=L$.
The choice \refb{echoiceint} 
of course is not unique since we can deform the contour
without changing the result for the index as long as we do not
cross a pole of the partition function.

Independent of the above analysis,
the change in the index across a wall of marginal
stability can be computed using the wall crossing 
formula\cite{0005049,0010222,0101135,0206072,
0304094,0702141,0702146}. This
tells us that
as we cross a wall of marginal stability
associated with the decay $(Q,P)\to (Q_1,P_1)+(Q_2,P_2)$, 
the index
jumps by an amount\footnote{Eq.\refb{eff3} holds only if the
dyons $(Q_1,P_1)$ and $(Q_2,P_2)$ are primitive. As will be
discussed later, this formula gets modified for non-primitive decay.}
\be \label{eff3}
(-1)^{Q_1\cdot P_2 - Q_2 \cdot P_1+1} \,
(Q_1\cdot P_2 - Q_2 \cdot P_1) \,
d_h(Q_1,P_1) d_h(Q_2, P_2)\, 
\ee
up to a sign,
where $d_h(Q,P)$ denotes the index of half-BPS states carrying
charge $(Q,P)$.  For the decay described in
\refb{eff1} the relevant half-BPS indices
are of the form $d_h(a_0 M_0, c_0 M_0)$ and 
$d_h( b_0 N_0, d_0 N_0)$ where $M_0\equiv d_0 Q - b_0 P$ and
$N_0\equiv -c_0 Q + a_0 P$. T-duality invariance implies that
-- modulo some subtleties discussed below eqs.\refb{ex5} --
the dependence of
$d_h(a_0 M_0, c_0 M_0)$ and 
$d_h( b_0 N_0, d_0 N_0)$ on $M_0$ and $N_0$ 
must come via the combinations
$M_0^2$ and $N_0^2$ respectively. 
We now define
\be \label{eff3.5}
\phi_e(\tau;a_0, c_0) \equiv \sum_{M_0^2} e^{\pi i\tau M_0^2} 
d_h(a_0 M_0, c_0 M_0), \quad
\phi_m(\tau;b_0, d_0) \equiv \sum_{N_0^2} e^{\pi i\tau N_0^2}
d_h( b_0 N_0, d_0 N_0)\, ,
\ee
where the sums are over the sets of $(M_0^2, N_0^2)$ values which
arise in the possible decays of the dyons in the
set $\BBB$ via \refb{eff1}. Then
\refb{eff3} agrees with the residue of the partition function at the
pole \refb{eff2} if we assume that $\cp$ has
a double zero at \refb{eff2} where it 
behaves as
\ben \label{eff4}
\cp(\crh,\cs,\cv) &\propto&
\cv^{\prime 2}\, \phi_e(\cs' ;a_0, c_0) \, 
\phi_m(\crh'; b_0, d_0)\, ,  \qquad 
\cv'\equiv \crh c_0d_0 + \cs a_0b_0 + \cv (a_0d_0 + b_0c_0),
\nonumber \\ &&
\cs'\equiv  c_0^2\crh + a_0^2
\cs  + 2 a_0c_0 \cv, \quad \crh' \equiv
d_0^2\crh + b_0^2 \cs 
+ 2 b_0d_0 \cv\, . 
\een  
Since for any given system the allowed values
of $(a_0,b_0,c_0,d_0)$ can be found from charge quantization laws,
\refb{eff4} gives us information about the locations of the zeroes
on $\cp$ and its behaviour at these zeroes in terms of the spectrum
of half-BPS states in the theory.

\item {\bf Additional modular symmetries:}
Often the partition functions associated with $d_h(Q,P)$ have
modular properties, {\it e.g.} the function $\phi_m(\tau;a_0,c_0)$
could  transform as a modular form under $\tau\to(\alpha\tau+\beta)
/(\gamma\tau+\delta)$
and $\phi_e(\tau;b_0,d_0)$ 
could transform as a modular form under 
$\tau\to (p\tau+q)/(r\tau+s)$ with $\pmatrix{\alpha&\beta\cr \gamma
&\delta}$ and $\pmatrix{p & q\cr r & s}$ belonging to  
certain subgroups
of $SL(2,\ZZZ)$.  Some of these may be accidental symmetries,
but some could be consequences of exact symmetries of the
full partition function $\cp(\crh,\cs,\cv)^{-1}$. Using
\refb{eff4} one finds 
 that those which can be lifted to exact symmetries of
$\cp$ can be represented as symplectic transformations of
the form \refb{ei6} with $\pmatrix{A & B\cr C & D}$ given by
\be \label{eff7}
\pmatrix{d_0 & b_0 &0&0 \cr c_0 & a_0 &0&0\cr
0&0& a_0 & -c_0\cr 0&0& -b_0 & d_0}^{-1}
\pmatrix{\alpha & 0 & \beta & 0\cr
0 & 1 & 0 & 0\cr \gamma & 0 & \delta & 0\cr
0 & 0 & 0 & 1}
\pmatrix{d_0 & b_0 &0&0 \cr c_0 & a_0 &0&0\cr
0&0& a_0 & -c_0\cr 0&0& -b_0 & d_0} 
\ee
and
\be \label{eff8}
\pmatrix{d_0 & b_0 &0&0 \cr c_0 & a_0 &0&0\cr
0&0& a_0 & -c_0\cr 0&0& -b_0 & d_0}^{-1}
\pmatrix{1 & 0 & 0 & 0\cr
0 & p & 0 & q\cr 0 & 0 & 1 & 0\cr
0 & r & 0 & s}
\pmatrix{d_0 & b_0 &0&0 \cr c_0 & a_0 &0&0\cr
0&0& a_0 & -c_0\cr 0&0& -b_0 & d_0}
\ee
respectively.
These represent additional
symmetries of $\cp$ besides the ones associated with S-duality
invariance and charge quantization laws. 
Furthermore the constant $k$ appearing in \refb{ei6} is
given by the weight of $\phi_e$ and $\phi_m$ minus 2. 

It is these additional symmetries which make the symmetry group
of $\cp$ a non-trivial subgroup of $Sp(2,\ZZZ)$. 
The
S-duality transformations \refb{ei7} and the translation symmetries
\refb{ei8} are both associated with $Sp(2,\ZZZ)$ matrices 
$\pmatrix{A & B\cr C & D}$ with $C=0$. In contrast the
trnsformations \refb{eff7}, \refb{eff8} typically have $C\ne 0$.

Since we do not {\it a priori} know which part of the modular
symmetries of $\phi_e$ and $\phi_m$ survive as symmetries of
$\cp$, this does not give a foolproof method for identifying
symmetries of $\cp$. However often by combining information
from the behaviour of $\cp$ around different zeroes one can make
a clever guess.

\item {\bf Black hole entropy:}
Additional constraints may be found be requiring that
in the limit of large charges the index reproduces correctly the
black hole entropy.\footnote{Here we are implicitly assuming that
when the effect of interactions are 
taken into account, only index worth of states
remain as BPS states so that the black hole entropy can be
compared to the logarithm of the index.}
In particular, by requiring that we reproduce the
black hole entropy $\pi\sqrt{Q^2P^2-(Q\cdot P)^2}$ that arises in
the supergravity approximation one finds that $\cp$ is required
to have a zero 
at\cite{9607026,0412287,0510147,0605210,0708.1270}
\be \label{ebb11}
\crh \cs - \cv^2 + \cv=0\, .
\ee
In order to find the behaviour of $\cp$ near this zero one needs to
calculate the first non-leading correction to the black hole
entropy and compare this with the first non-leading correction to
the formula for the index. In general the former requires the
knowledge of the complete set of four derivative terms in the effective
action, but in all known examples one can reproduce the answer for
the index just by taking into account
the effect of the Gauss-Bonnet term in the action.
If we assume that this continues to hold in general then by matching
the first non-leading corrections on both sides one can relate the
behaviour of $\cp$ near \refb{ebb11} 
to the coefficient of the Gauss-Bonnet
term. The result is
\be \label{ebb2}
\cp(\crh, \cs,\cv) \propto (2v-\rho-\sigma)^k\,
\{ v^2 \, g(\rho)\, g(\sigma) +  \OO(v^4)\}\, ,
\ee
where
\be \label{ebb3}
\rho = {\crh \cs - \cv^2\over \cs}, \quad
\sigma = {\crh \cs - (\cv-1)^2\over \cs}, \quad
v= {\crh \cs - \cv^2+\cv\over \cs}\, ,
\ee
and $g(\tau)$ is a modular form of weight $k+2$ of the S-duality
group, related to the Gauss-Bonnet term
\be \label{ebb4}
 \int d^4 x\, \sqrt{-\det g} \,
 \phi(a,S)\, 
\left\{ R_{\mu\nu\rho\sigma} R^{\mu\nu\rho\sigma}
- 4 R_{\mu\nu} R^{\mu\nu}
+ R^2
\right\} \, ,
\ee
via the relation
\be \label{ebb5}
\phi(a,S) = - {1\over 64\pi^2} \, \left( (k+2) \ln S 
+ \ln g(a+iS) + \ln g(-a+iS)\right)
+\hbox{constant}\, .
\ee
Here $\tau=a+iS$ is the axion-dilaton modulus.

\end{enumerate}

In 
\S\ref{s3} 
we apply the considerations 
described above to several examples.
These include known examples involving unit torsion dyons
in heteroric string theory on $T^6$ and CHL orbifolds and also
some unknown cases like dyons of torsion 2 in heterotic string
theory on $T^6$  
(\i.e.\ dyons for which
gcd($Q\wedge P$)=2\cite{0702150})
and dyons carrying untwisted sector charges
in $\ZZZ_2$ CHL orbifold\cite{9505054,9506048}. 
In the latter cases we
determine the constraints imposed by the S-duality invariance
and wall crossing formul\ae\ and also try to use the known
modular properties of half-BPS states to guess the symmetry
group of the quarter BPS dyon partition function.

In \S\ref{sprop} we propose a formula for the dyon partitions
function of  torsion two dyons in heterotic string theory on $T^6$.
The formula for the partition function when $Q$ and $P$ are
both primitive but $(Q\pm P)$ are twice primitive vectors is
\ben \label{etor2}
{1\over \cp(\crh,\cs,\cv)} &=& {1\over 8} \Bigg[
{1\over \Phi_{10}(\crh,\cs,\cv)} +
{1\over \Phi_{10}(\crh+{1\over 2},\cs+{1\over 2},\cv)} +
{1\over \Phi_{10}(\crh+{1\over 4},\cs+{1\over 4},\cv+{1\over 4})} \nonumber \\
&&+
{1\over \Phi_{10}(\crh+{3\over 4},\cs+{3\over 4},\cv+{1\over 4})} 
 +
{1\over \Phi_{10}(\crh+{1\over 2},\cs+{1\over 2},\cv+{1\over 2})} \nonumber \\ 
&& +
{1\over \Phi_{10}(\crh,\cs,\cv+{1\over 2})}  +
{1\over \Phi_{10}(\crh+{3\over 4},\cs+{3\over 4},\cv+{3\over 4})} +
{1\over \Phi_{10}(\crh+{1\over 4},\cs+{1\over 4},\cv+{3\over 4})}\Bigg]
\nonumber \\  
&&  + 
{2\over \Phi_{10}(\crh+\cs+2\cv,
\crh+\cs-2\cv,\cs-\crh)}
\een
where $\Phi_{10}$ is the weight 10
Igusa cusp form of $Sp(2,\ZZZ)$ describing the inverse
partition function of torsion one dyons. The sum of the first eight
terms on the right hand side of \refb{etor2}
coincides with the partition function of unit torsion dyons
subject to the constraints that $Q^2+P^2\pm 2Q\cdot P$ are
multiples of 8; the last term is a new addition.
We show that \refb{etor2} satisfies all the required
consistency conditions. First of all it has the required S-duality
invariance. It also satisfies the wall crossing formul\ae\ at all
the walls of marginal stability at which the original dyon decays
into a pair of primitive dyons.
It satisfies the constraint \refb{ebb2} coming from the requirement
that the statistical entropy and the black hole entropy agrees
up to the first non-leading order in inverse powers of charges.
Furthermore by taking an appropriate
limit of this formula we can reproduce the known results for
torsion two dyons in gauge 
theories\cite{9804174,9907090,0005275,0609055,appear}.

In the case of torsion two dyons with $Q$, $P$ both
primitive,  the vectors $Q\pm P$ are not primitive, but 
$(Q\pm P)/2$ are
primitive vectors\cite{0801.0149}. 
As a result for the decay into
\be \label{eq1p1}
(Q_1,P_1) = (Q-P,0), \qquad (Q_2,P_2) = (P,P)\, ,
\ee
the charge vector $(Q_1,P_1)$ is not primitive. Computing the
jump in the index from \refb{etor2} we find that in this case
the change in the index across this wall of marginal stability is
given by
\be \label{efh7int}
\Delta d(Q,P) = (-1)^{Q_1\cdot P_2 - Q_2\cdot P_1 + 1}
(Q_1\cdot P_2 - Q_2\cdot P_1)\, 
\left\{d_h\left(Q_1,P_1\right) +
d_h\left({1\over 2} Q_1,{1\over 2} P_1\right) \right\}
\, d_h\left(Q_2,P_2\right) \, .
\ee
This differs from the formula \refb{eff3}. A similar modification of
the wall crossing formula for decays into non-primitive states in
$\NN=2$ supersymmetric string theories has been suggested in
\cite{0702146}.

There are two more classes of dyons of torsion two, -- one where
$Q$ is primitive and $P$ is twice a primitive vector and the other
where $P$ is primitive and $Q$ is twice a primitive vector. The
partition functions for these dyons can be recovered from the
one given above by S-duality transformations $(Q,P)\to (Q, P-Q)$
and $(Q,P)\to (Q-P, P)$ respectively\cite{0801.0149}. This amount
to making replacements
$(\crh,\cs,\cv)\to(\crh,\cs+\crh+2\cv,\cv+\crh)$ and
$(\crh,\cs,\cv)\to(\crh+\cs+2\cv,\cs,\cv+\cs)$ 
respectively in eq.\refb{etor2}.

Our analysis can also be used to predict the form of the partition
function of dyons of higher torsion. These results will be presented
in a forthcoming publication\cite{rforth}.

Although we have presented most of our analysis as a way of
extracting information about the partition function of quarter
BPS states from known spectrum of half-BPS states, we could
also use it in the reverse direction. 
In the final section \S\ref{sapp} we provide some examples in
the context of $\ZZZ_N$ orbifold models 
where the knowledge of the
quarter BPS partition function can be used to compute the
spectrum of a certain class of half-BPS states.

\sectiono{The dyon partition function}
\label{s2}

Let us consider a particular $\NN=4$ supersymmetric string
theory in four dimensions with a total of $r$ U(1) gauge fields
including the six graviphotons. 
The electric and magnetic charges in this theory are represented
by $r$ dimensional vectors $Q$ and $P$, and there is a T-duality
invariant metric $L$ of signature $(6,r-6)$ that can be used to
define the inner product of the charges.
Let us consider an (infinite) set $\BBB$ of dyon charge vectors
$(Q,P)$ with the property that 
if two different members of the set have the same values
of $Q^2\equiv Q^T L Q$, $P^2\equiv P^T L P$ and 
$Q\cdot P\equiv Q^T L P$ then there must exist a
T-duality transformation that relates the two 
members.
In other words
if there are  T-duality invariants other than $Q^2$, $P^2$ and
$Q\cdot P$ then for all members of the set $\BBB$ with a given
set of values of $(Q^2, P^2, Q\cdot P)$
these other T-duality invariants
must have the same values. 
We shall generate such a set $\BBB$ by  
beginning with a family $\AAA$ of charge vectors 
$(Q,P)$ labelled by three integers such that the 
triplet
$(Q^2,P^2,Q\cdot P)$ are independent linear functions of these
three integers, and then
define $\BBB$ to be the set of all $(Q,P)$ which are in the
T-duality orbit of the set $\AAA$. Such a set $\BBB$ automatically
satisfies the restriction mentioned above since given two elements
of $\BBB$ with the same values of $(Q^2,P^2,Q\cdot P)$, each will be
related by a T-duality transformation to the unique element
of $\AAA$ with these values of $(Q^2,P^2,Q\cdot P)$.
An example of such a set $\AAA$ can be found in eqs.\refb{ess3},
\refb{ess4}.

Our object of interest is the index $d(Q,P)$, measuring the number
of bosonic supermultiplets minus the number of fermionic supermultiplets
of quarter BPS dyons
carrying charges $(Q,P)\in \BBB$.
Typically the index, besides depending on $(Q,P)$,
also depends of the domain in which
the asymptotic moduli lie.
These domains are bounded by walls of marginal stability associated
with decays of the form $(Q,P)\to 
(\alpha Q + \beta P, \gamma Q+\delta P) + 
((1-\alpha) Q - \beta P, -\gamma Q+(1-\delta) P)$ for
appropriate values of $(\alpha,\beta,\gamma,\delta)$ associated
with the quantization 
conditions\cite{0707.1563,0707.3035,0710.4533}.
For fixed values of the other moduli these walls describe circles
or straight lines in the axion-dilaton moduli space labelled
by the complex parameter $\tau$\cite{0702141,0707.1563}.
We denote by 
$\vec c$ the collection of $(\alpha_i, \beta_i, \gamma_i,\delta_i)$
bordering a particular domain in the moduli space; inside any such
domain the  index remains unchanged. 
It has been shown in \cite{0702141} that 
the parameters $\vec c$ labelling a
domain remain invariant under a simultaneous T-duality
transformation on the charges and the moduli. Since $d(Q,P)$ must
be invariant under simultaneous T-duality transformation on the
charges and the moduli, we can conclude that for a given
$\vec c$  the index $d(Q,P)$ for $(Q,P)\in\BBB$ 
will be a function only of the
T-duality invariants $(Q^2,P^2,Q\cdot P)$. We shall express this
as 
$f(Q^2, P^2, Q\cdot P, \vec c)$.

Let us now introduce the partition function
\be \label{e1}
{1\over \cp(\crh, \cs, \cv)}
\equiv \sum_{Q^2, P^2, Q\cdot P}\, 
(-1)^{Q\cdot P+1}\, f(Q^2,P^2, Q\cdot P; \vec c_0) 
e^{i\pi (\cs Q^2 + \crh P^2 
+ 2 \cv Q\cdot P)}\, .
\ee
where $\vec c_0$ denotes some specific domain in the moduli
space bounded by a set of walls of marginal stability.
The sum runs over allowed values of $Q^2$, $P^2$ and $Q\cdot P$
for the dyons belonging to the set $\BBB$. 
The factor of $(-1)^{Q\cdot P+1}$ has
been included for convenience.
$\cp$ so defined is expected to be a periodic
function of $\crh$, $\cs$ and $\cv$, with the
periods depending on the
quantization condition on $P^2$, $Q^2$ and $Q\cdot P$.
Let the periods be $T_1$, $T_2$ and $T_3$ respectively
-- these represent inverses of the quanta of $P^2/2$, $Q^2/2$
and $Q\cdot P$ belonging to the set $\BBB$. 
The sum given in \refb{e1} is typically not convergent
for real values of $\crh$, $\cs$ and $\cv$. However
often it may be made convergent by treating $\crh$, 
$\cs$ and $\cv$ as complex variables and working
in appropriate domain in the complex plane.
We shall assume that this can be done.
We may now invert \refb{e1} as
\ben \label{e2}
f(Q^2,P^2, Q\cdot P; \vec c_0)
&=& {(-1)^{Q\cdot P+1}\over T_1 T_2 T_3} 
\int_{i M_1-T_1/2}^{iM_1+T_1/2} d\crh
\int_{iM_2-T_2/2}^{i M_2+T_2/2} d\cs 
\int_{i M_3-T_3/2}^{i M_3+T_3/2} d\cv \, \nonumber \\
&& \qquad \qquad
e^{-i\pi (\cs Q^2 + \crh P^2 
+ 2 \cv Q\cdot P)} \, 
{1\over \cp(\crh, \cs, \cv)}\, ,
\een
provided the imaginary parts $M_1$, $M_2$ and $M_3$ of 
$\crh$, $\cs$ and $\cv$ are fixed in a region where the
original sum \refb{e1} is convergent.

During the above discussion we have implicitly assumed that
the quantization laws of $Q^2$, $P^2$ and $Q\cdot P$ are
uncorrelated so that $\cp(\crh,\cs,\cv)$ is separately invariant under
$\crh\to\crh+T_1$, $\cs\to\cs+T_2$ and $\cv\to \cv+T_3$.
In general we can have more complicated periods which involve
simultaneous shifts of $\crh$, $\cs$ and $\cv$. In this case the
integration in \refb{e2} needs to be carried out over an
appropriate unit cell in the $(\Re(\crh),\Re(\cs),\Re(\cv))$ space
and the factor of $T_1T_2T_3$ in the denominator will be
replaced by the volume of the unit cell.

\sectiono{Consequences of S-duality symmetry} \label{sdual}

Let us now assume that the theory has S-duality symmetries
of the form
\be \label{e3}
Q\to Q'' = a Q + b P, \qquad P\to P''=c Q + d P\, ,
\ee
for appropriate choice of $(a,b,c,d)$. Under this transformation
\be \label{e3.5}
Q^{\ppp 2} = a^2 Q^2 + b^2 P^2 + 2 ab Q\cdot P,
\quad P^{\ppp 2} = c^2 Q^2 + d^2 P^2 + 2 cd Q\cdot P,
\quad Q''\cdot P'' = ac  Q^2 + bd  P^2 + (ad+bc) Q\cdot P\, .
\ee
A generic S-duality transformation acting on an arbitrary
element of $\BBB$
will give rise to $(Q'',P'')$ outside the set $\BBB$ for which
the index formula is given by the function $f$. 
We shall restrict ourselves
to a subset of S-duality transformations which takes an element
of the set $\BBB$ to another element of the set $\BBB$.  
For such transformations, the S-duality
invariance of the theory tells us that
\be \label{e4}
f(Q^{\ppp 2}, P^{\ppp 2}, Q''\cdot P'', \vec c\,''_0)
= f(Q^2, P^2, Q\cdot P, \vec c_0)\, ,
\ee
where $\vec c\,''_0$ denotes the collection 
$\{(\alpha_i'', \beta_i'', \gamma_i'',\delta_i'')\}$ of domain walls
related to the set
$\{(\alpha_i, \beta_i, \gamma_i,\delta_i)\}$ associated
with $\vec c_0$ by the relation\cite{0702141,0801.0149}
\be \label{e5}
\pmatrix{ \alpha_i'' & \beta_i''\cr  \gamma_i'' & \delta_i''}
= \pmatrix{a & b\cr c & d} \pmatrix{ \alpha_i & \beta_i
\cr  \gamma_i & \delta_i} \pmatrix{a & b\cr c & d}^{-1}\, .
\ee
Physically the domain corresponding to $\vec c\,''_0$ represents
the image of the one corresponding to $\vec c_0$ under 
simultaneous S-duality transformation on the charges and the
moduli.
Making a change of variables
\be \label{e6}
\crh = d^2 \crh'' + b^2 \cs'' + 2 bd \cv'', 
\quad \cs = c^2 \crh'' + a^2 \cs'' + 2 ac \cv'',
\quad \cv = cd \crh'' + ab \cs'' + (ad+bc) \cv''\, ,
\ee
in \refb{e2} and using the fact that 
\ben \label{efacts}
&& (-1)^{Q\cdot P} = (-1)^{Q''\cdot
P''}, \quad \cs Q^2 +\crh P^2 + 2 \cv Q\cdot P
=  \cs'' Q^{\prime\prime 2} +\crh'' P^{\prime\prime 2} 
+ 2 \cv'' Q''\cdot P'', \nonumber \\ && \qquad
d\crh\wedge d\cs\wedge d\cv = d\crh''\wedge d\cs''\wedge d\cv''\, ,
\een
under an S-duality transformation,
we can express \refb{e4} as
\be \label{e7}
f(Q^{\ppp 2}, P^{\ppp 2}, Q''\cdot P'', \vec c\,''_0)
= {(-1)^{Q''\cdot P''+1}\over T_1 T_2 T_3}\, 
\int_\CC \, d\crh''
d\cs''  d\cv'' \,
e^{-i\pi (\cs'' Q^{\ppp 2} + \crh'' P^{\ppp 2} 
+ 2 \cv'' Q''\cdot P'')} \, 
{1\over \cp(\crh, \cs, \cv)}\, ,
\ee
where $\CC$ is the image of the original region of integration
\refb{e2} in the complex $(\crh'', \cs'', \cv'')$ plane:
\ben \label{econtour}
&& \Im(\crh'') = a^2 M_1 + b^2 M_2 - 2 ab M_3, \quad
\Im(\cs'') = c^2 M_1 + d^2 M_2 - 2 cd M_3, \nonumber \\
&&
\Im (\cv'') = - ac M_1 - bd M_2 + (ad + bc) M_3\, .
\een

We would like to get some constraint on the function $\cp$
by comparing \refb{e2} with \refb{e7}. For this we note that
we can replace $(Q,P)$ by $(Q'', P'')$ and 
$(\crh,\cs, \cv)$ by $(\crh'', \cs'', \cv'')$ everywhere
in \refb{e2} since they are dummy variables. This gives
\ben \label{e8}
f(Q^{\ppp 2},P^{\ppp 2}, 
Q''\cdot P''; \vec c_0)
&=& {(-1)^{Q''\cdot P''+1}\over T_1 T_2 T_3} 
\int_{iM_1-T_1/2}^{iM_1+T_1/2} d\crh''
\int_{iM_2-T_2/2}^{iM_2+T_2/2} d\cs'' 
\int_{iM_3-T_3/2}^{iM_3+T_3/2} d\cv'' \,
 \nonumber \\ &&  \qquad \qquad \qquad
e^{-i\pi (\cs'' Q^{\ppp 2} + \crh'' P^{\ppp 2}
+ 2 \cv'' Q''\cdot P'')} 
{1\over \cp(\crh'', \cs'', \cv'')}\, .
\een
Since 
in general $\vec c_0$ and $\vec c\,''_0$ describe different
domains, we cannot compare \refb{e7} and 
\refb{e8} to constrain the
form of $\cp$ without any further input.\footnote{The only
exceptions are those S-duality transformations which leave the
domain $\vec c_0$ unchanged\cite{0702141}.} 
However the dyon spectrum in
a variety of $\NN=4$ supersymmetric string theories displays
the feature that the spectrum in two different domains 
$\vec c\,''_0$ and $\vec c_0$ are both given as integrals with the
same integrand, but for $\vec c\,''_0$
the integration over $(\crh'',\cs'',
\cv'')$ is carried out over a different subspace  
 than the one given
in \refb{e8}.  In particular if $\vec c\, ''_0$ is related to
$\vec c_0$ by an S-duality transformation then this subspace is
given by the integration region $\CC$ given in \refb{econtour}.
We shall assume that this feature continues to
hold in the general situation. In that case the effect of
replacing
$\vec c_0$ by $\vec c\, ''_0$ in eq.\refb{e8} is
to replace the integration contour 
by $\CC$ on the right hand side:
\be \label{e9}
f(Q^{\ppp 2},P^{\ppp 2}, 
Q''\cdot P''; \vec c\,''_0)
= {(-1)^{Q''\cdot P''+1}\over T_1 T_2 T_3} \int_\CC d\crh''
d\cs'' d\cv'' \,
e^{-i\pi (\cs'' Q^{\ppp 2} + \crh'' P^{\ppp 2}
+ 2 \cv'' Q''\cdot P'')} \, 
{1\over \cp(\crh'', \cs'', \cv'')}\, .
\ee
Comparing \refb{e9} and \refb{e7} we get
\be \label{e10}
\cp(\crh, \cs, \cv)= \cp(\crh'', \cs'', \cv'') \, .
\ee

For future reference we shall rewrite the transformation laws
\refb{e6} in a suggestive form. We define
\be \label{e11}
\com = \pmatrix{\crh & \cv\cr \cv & \cs}\, .
\ee
Then the transformations \refb{e6} may be written as
\be \label{e12}
\com = (A\com'' + B) (C\com'' + D)^{-1}
\, ,
\ee 
where $A$, $B$, $C$ and $D$ are $2\times 2$ matrices, given by
\be \label{e13}
\pmatrix{A & B\cr C & D} = \pmatrix{d & b &0&0 \cr c & a &0&0\cr
0&0& a & -c\cr 0&0& -b & d}\, .
\ee
Eq.\refb{e10} now gives (after replacing the dummy variable
$\com''$ by $\com$ on both sides),
\be \label{espec}
\cp((A\com + B) (C\com + D)^{-1})
= \det(C\com + D)^k \cp(\com)\, ,
\ee
for $A$, $B$, $C$, $D$ given in \refb{e13}.
Here $k$ is an arbitrary number. Since $\det(C\Omega +D)=1$,
we cannot yet ascertain the value of $k$.

To this we can also append the translational symmetries of
$\cp$:
\be \label{e14}
\cp(\crh, \cs, \cv) =
\cp(\crh+a_1, \cs+a_2, \cv+a_3)\, ,
\ee
where $a_i$'s are integer multiples of the $T_i$'s. It is
convenient, although not necessary, to work with appropriately
rescaled $Q$ and/or $P$ so that the 
$T_i$'s and hence the 
$a_i$'s are integers. 
This symmetry can also be rewritten as \refb{espec} with the choice
\be \label{e15}
\pmatrix{A & B\cr C & D} = \pmatrix{1 & 0 & a_1 & a_3\cr
0 & 1 & a_3 & a_2\cr 0 & 0 & 1 & 0\cr 0 & 0 & 0 & 1}\, .
\ee
Again since $\det(C\com+D)=1$ the choice of $k$ is
arbitrary.

The alert reader would have noticed that although we have expressed
the consequences of S-duality invariance and charge quantization
conditions as symmetries of the function
$\cp$ under a symplectic transformation, 
the symplectic transformations arising this way are
trivial, --  for all the transformations arising this way the matrix
$C$ vanishes and hence the transformations act linearly on the
variables $\crh$, $\cs$ and $\cv$. In order to show that
the function $\cp$ has non-trivial modular properties we need
to find symmetries of $\cp$ which have non-vanishing $C$.
This will also determine the weight of $\cp$ under the modular
transformation. To get a hint about any possible
additional symmetries of $\cp$ we need
to make use of the wall crossing formula for the dyon spectrum
of $\NN=4$ supersymmetric string theories.
This will be the subject of discussion in \S\ref{swall}.

\sectiono{Constraints from wall crossing} \label{swall}

As has already been discussed, the index associated with the
quarter BPS
dyon spectrum in $\NN=4$ supersymmetric string theories can
undergo discontinuous jumps across walls of marginal stability.
{\it A priori} the formula for the dyon
spectrum in different domains labelled by the vector $\vec c$
could be completely different.
However the study of dyon spectrum in a variety
of $\NN=4$ supersymmetric string theories shows that
in different domains the index continues to be given by
an expression similar to
\refb{e2}, the only difference being that the choice of the
3 real dimensional subspace (contour) over 
which we carry out the integration
in the complex $(\crh,\cs, \cv)$ plane is different in
different domains. As a result the difference between the indices
in two different domains is given by the sum of residues
of the integrand at the poles we encounter while deforming the
contour associated with one domain to the contour associated with
another domain. 
As a special example of this we can consider the decay 
$(Q,P)\to (Q,0)+ (0,P)$.
In all known examples change in the index across this wall
of marginal stability is 
accounted for by the residue of a double 
pole of the integrand at $\cv=0$, \i.e.
as we cross this particular wall of marginal stability
in the moduli space, the integration contour crosses the pole at
$\cv=0$. Since the change in the index as we cross a given wall
can be found using the wall crossing 
formula\cite{0005049,0010222,0101135,0206072,
0304094,0702141,0702146}, this provides
information on the residue of the integrand at the $\cv=0$ pole.

There are many other possible
decays of a quarter BPS state into a pair of 
half BPS states.
All such decays may be parametrized as\cite{0702141}
\be \label{ede1}
(Q,P) \to (a_0d_0 Q - a_0b_0 P, c_0d_0 Q - c_0b_0 P) +
(-b_0c_0 Q + a_0b_0 P, - c_0d_0 Q + a_0d_0 P)\, , \quad a_0
d_0 - b_0c_0 =1\, .
\ee
$a_0$, $b_0$, 
$c_0$, $d_0$ are not necessarily all integers, but must be
such that the charges carried by the decay products belong to
the charge lattice. One can try 
to use the wall crossing formul\ae\ associated with these
decays to further constrain the form of $\cp$.
For unit torsion states in heterotic string theory on $T^6$, $a_0$, 
$b_0$,
$c_0$ and $d_0$ are integers and the decay given in \refb{ede1} is
related to the decay $(Q,P)\to (Q,0)+(0,P)$ via an S-duality
transformation $\pmatrix{a_0 & b_0\cr c_0 & d_0}$.
Thus the change in the index across the wall
is controlled by the residue of the partition function at 
a new pole that is
related to the $\cv=0$ pole by the S-duality transformation
\refb{e6}. This gives the location of the pole to be at
\be \label{epole1}
\crh c_0d_0 + \cs a_0b_0 + \cv (a_0d_0 + b_0c_0) = 0\, .
\ee
As long as $\cp$ is manifestly S-duality invariant, \i.e.\
satisfies \refb{e13}, \refb{espec}, the residues at these poles will
automatically satisfy the wall crossing formula. Thus they do not
provide any new information. However
in a generic situation new walls may appear,  labelled by
fractional values of $a_0,b_0,c_0,d_0$. Also
the S-duality group is smaller. As a result not all
the walls can be related to each other by S-duality transformation.
It is tempting to speculate that the jump across
any wall of marginal stability associated with the decay
\refb{ede1} is described by the residue of the partition function
at the pole at \refb{epole1}. We shall proceed with this
assumption -- this will be one of our key 
postulates.\footnote{Of course the translation symmetries
\refb{e14} allow us to shift a pole at \refb{epole1} to other
equivalent locations. Our postulate asserts that the contribution comes
from poles which can be brought to \refb{epole1} using the
translation symmetries \refb{e14}.  In that case we can choose the
unit cell over which we carry out the 
integration  in \refb{e2} in such a way that only the pole at 
\refb{epole1} contributes to the jump in the index across
the wall at \refb{ede1}.
A possible
exception to this will be discussed in the paragraphs
above eq.\refb{esx1} 
where we address some subtle issues.}

Before we proceed we shall show that this postulate is internally
consistent, \i.e.\ it is possible to choose $\CC$ at different points
in the moduli space consistent with this postulate. For this we
generalize the contour prescription of \cite{0706.2363}, assuming
that it holds for all $\NN=4$ supersymmetric string theories. Let
$\tau=\tau_1+i\tau_2$ be the axion-dilaton moduli, $M$ be the
usual $r\times r$ symmetric
matrix valued moduli satisfying $MLM^T=L$, and
\be \label{edefcr}
Q_R^2 \equiv Q^T (M+L)Q, \quad P_R^2\equiv
P^T (M+L)P, \quad
Q_R\cdot P_R \equiv Q^T(M+L)P\, .
\ee
Then at the point $(\tau,M)$ in the space of asymptotic moduli
we choose $\CC$ to be
\ben \label{echoice}
\Im(\crh) &=& \Lambda \, \left({|\tau|^2\over \tau_2} +
{Q_R^2 \over \sqrt{Q_R^2 P_R^2 - (Q_R\cdot P_R)^2}}\right)\, ,
\nonumber \\
\Im(\cs) &=& \Lambda \, \left({1\over \tau_2} +
{P_R^2 \over \sqrt{Q_R^2 P_R^2 - (Q_R\cdot P_R)^2}}\right)\, ,
\nonumber \\
\Im(\cv) &=& -\Lambda \, \left({\tau_1\over \tau_2} +
{Q_R\cdot P_R \over \sqrt{Q_R^2 P_R^2 - (Q_R\cdot P_R)^2}}
\right)\, ,
\een
where $\Lambda$ is a large positive number. Then on $\CC$
\ben \label{emar1}
&&\Im( c_0d_0\crh + a_0 b_0 \cs +(a_0d_0+b_0c_0)\cv)
\nonumber \\
&=&{c_0 d_0 \over \tau_2}\, \Lambda\, \left\{
\left(\tau_2+{E\over 2 c_0 d_0}\right)^2 +\left(
\tau_1 - {a_0d_0+b_0c_0\over 2c_0d_0}\right)^2 - \left(1
+{E^2\over 4c_0^2d_0^2}\right)\right\}\, ,
\een
where
\be \label{emar2}
E={c_0 d_0 Q_R^2 +a_0b_0 P_R^2 - (a_0d_0+b_0c_0)Q_R\cdot P_R
\over \sqrt{Q_R^2P_R^2-(Q_R\cdot P_R)^2}}\, .
\ee
As shown in \cite{0702141}, the right hand side of \refb{emar1}
vanishes on the wall of marginal stability associated with the
decay  given in \refb{ede1}. Thus it follows from \refb{emar1} that
as we cross this wall of marginal stability, the contour
\refb{echoice} crosses the pole at \refb{epole1} in accordance with
our postulate.

This postulate allows us to identify the
possible poles of the partition function besides those related
to the $\cv=0$ pole by the S-duality transformation
\refb{e6}, -- they occur at
\refb{epole1} for those values of 
$a_0$, $b_0$, $c_0$ and $d_0$ for which
the decay \refb{ede1} is consistent
with the charge quantization laws. 
One can also get information about the residues at these poles
since they are given by the jumps in the index.
This jump can be expressed
using the wall crossing 
formula\cite{0005049,0010222,0101135,0206072,
0304094,0702141,0702146} 
that tells us that
as we cross a wall of marginal stability
associated with the decay $(Q,P)\to (Q_1,P_1)+(Q_2,P_2)$ the index
jumps by an amount
\be \label{ex1}
(-1)^{Q_1\cdot P_2 - Q_2 \cdot P_1+1} 
\, (Q_1\cdot P_2 - Q_2 \cdot P_1)\, 
d_h(Q_1,P_1) d_h(Q_2, P_2)\, 
\ee
up to a sign,
where $d_h(Q,P)$ denotes the index measuring the number
of bosonic minus the number of fermionic half BPS 
supermultiplets carrying
charges $(Q,P)$. Thus this relates the residues at
the poles of the integrand to the indices of half BPS states.

We shall now study the consequence of \refb{ex1} 
on the residue at the pole \refb{epole1}. 
First let us consider the special case associated with 
the decay 
$(Q,P)\to (Q,0)+ (0,P)$. In this case the jump in the index
is given by
\be \label{ex2}
(-1)^{Q\cdot P+1} \, Q\cdot P \, d_e(Q) \, d_m(P)\, ,
\ee
where $d_e(Q)= d_h(Q,0)$ is the index of purely electrically
charged states and $d_m(P)=d_h(0,P)$ is the index of purely
magnetically charged state. This jump is to be
accounted for by the residue of a 
pole of the integrand at $\cv=0$.
The result \refb{ex2} is reproduced 
if near $\cv=0$, $\cp$ behaves as
\be \label{ex3}
\cp(\crh, \cs, \cv)^{-1} \propto \{
\phi_m(\crh)^{-1}\,  \phi_e(\cs)^{-1} \cv^{-2} + \OO(\cv^0)\}\, ,
\ee
where $1/\phi_m(\crh)$ and $1/\phi_e(\cs)$ denote respectively
the partition functions of purely
magnetic and purely
electric states:
\be \label{ex4}
d_m(P) = {1\over T_1}\,
\int_{iM_1-T_1/2}^{iM_1+T_1/2} 
d\crh\, e^{-i\pi P^2\crh} {1\over \phi_m(\crh)}\, ,
\qquad
d_e(Q) = {1\over T_2}\,
\int_{iM_2-T_2/2}^{iM_2+T_2/2} d\cs\, e^{-i\pi Q^2\cs} {1\over 
\phi_e(\cs)}\, .
\ee
Substituting \refb{ex4} into the integrand in \refb{e2} and
picking up the residue from the pole at $\cv=0$ we get the change
in the index to be
\be \label{ex5}
(-1)^{Q\cdot P+1} \, Q\cdot P \, d_e(Q) \, d_m(P)\, ,
\ee
in agreement with \refb{ex2}, provided we choose the constant of
proportionality in \refb{ex3} appropriately. Note that the
$Q\cdot P$ factor comes from the $\cv$ derivative of the
exponential factor in \refb{e2} arising due to the double pole
of $\cp^{-1}$ at $\cv=0$.

In writing \refb{ex3}, \refb{ex4} we have implicitly
assumed that the
allowed values of $Q^2$ and $P^2$ inside the set $\BBB$ 
are independent
of each other,
\i.e.\ the possible values that $Q^2$ can take for a given $P^2$ is
independent of $P^2$ and vice versa. 
If this is not so then instead of having a 
single product the right hand side of \refb{ex3} will contain
a sum of products. For example if in the set $\BBB$, $Q^2/2$ and
$P^2/2$ are correlated so that $Q^2/2$ is odd (even) when $P^2/2$
is odd (even) then the coeffcient of $\cv^{-2}$ in the expression
for $\cp^{-1}$ will contain two terms,
-- the product of the partition function with odd $Q^2/2$ electric
states with that of odd $P^2/2$ magnetic states and the product
of the partition function of even $Q^2/2$ electric states with that of
even $P^2/2$ magnetic states. 

There is one more assumption that has gone into writing
\refb{ex3}, \refb{ex4}. We have assumed that given two pairs
of charge vectors $(Q,P)$ and $(\wh Q,\wh P)$ in $\BBB$, 
if $Q^2
= \wh Q^{2}$ then $Q$ and $\wh Q$ are related
by a T-duality transformation. Otherwise $d_e(Q)$ will not
be  a function of $Q^2$ and one cannot define an electric partition
function via eq.\refb{ex4}. A similar restriction applies to
the magnetic charges as well. Now since  the set $\BBB$
has been chosen such that if the triplets $(Q^2,P^2,Q\cdot P)$ are
identical for two charge vectors then they must be related by
T-duality transformation, if two different $Q$'s with same
$Q^2$ are not related by T-duality then they must come from
triplets with different values of $P^2$ and/or $Q\cdot P$. In other
words the different T-duality orbits for a given $Q^2$ 
must be correlated with $P^2$ and/or 
$Q\cdot P$. If the correlation is with $P^2$ then we follow the
procedure described in the previous paragraph, {\it e.g.}
if one set of $Q$'s arise from even $P^2/2$ and another set
of $Q$'s arise from odd $P^2/2$, we define two separate electric
partition function for these two different sets of $Q$'s and
identify the coefficient of $\cv^{-2}$ in the partition
function $\cp^{-1}$ as a sum of terms.
If on the other hand the correlation is with $Q\cdot P$ then the
procedure is more complicated. We first project onto different
$Q\cdot P$ sectors by adding to $\cp^{-1}$ other terms obtained
by appropriate shifts of $\cv$, so that the subset of states
which contribute to the new partition function now has a
unique $Q$ for a given $Q^2$ up to T-duality transformations.
The singularities of this new partition functions near $\cv=0$ will now
be described by equation of the type \refb{ex3}, \refb{ex4}. For
example if one set of $Q$'s come from odd $Q\cdot P$ and the
second set of $Q$'s come from even 
$Q\cdot P$, then we can consider the quarter BPS partition functions
${1\over 2}\{
\cp^{-1}(\crh,\cs,\cv) \pm \cp^{-1}(\crh, \cs, \cv+{1\over 2})\}$.
These pick up even $Q\cdot P$ and odd $Q\cdot P$ states
respectively, and hence the contribution to these partition
functions will come from charge vectors $(Q,P)$ with the property
that for a given $Q^2$, there will be a unique $Q$ up to a
T-duality transformation. Thus the behaviour of these 
combinations will now be controlled by equations of the type
given in \refb{ex3}, \refb{ex4}. 
Conversely, for the original set $\BBB$ the
jump in the index associated with the decay $(Q,P)\to (Q,0)+(0,P)$
is now controlled by the zeroes of $\cp(\crh,\cs,\cv)$ at
$\cv=0$ and also at $\cv = 1/2$.
Similar considerations apply
when the same $P^2$ in the set $\BBB$ comes from more than
one $P$'s which are not related by T-duality. 

Often both the subtleties mentioned above can be avoided by a 
judicious choice of the set $\BBB$. In fact in all the explicit
examples we shall study in \S\ref{s3}, we shall be able to avoid
these subtleties.

We now return to the general case associated with the decay
described in \refb{ede1}.
Since 
here 
\be \label{esx1}
(Q_1,P_1)=(a_0d_0 Q - a_0b_0 P, c_0d_0 Q - c_0b_0 P), 
\qquad (Q_2,P_2)=
(-b_0c_0 Q + a_0b_0 P, - c_0d_0 Q + a_0d_0 P)\, ,
\ee
we have
\be \label{esx2}
Q_1\cdot P_2 - Q_2 \cdot P_1 = - Q^2 c_0d_0 - P^2 a_0b_0 
+ Q\cdot P
(a_0d_0 + b_0c_0)\, .
\ee
Let us now make a change of variables
\be \label{esx3}
\crh' = d_0^2\crh + b_0^2 \cs + 2 b_0d_0 \cv, 
\qquad \cs' = c_0^2\crh + a_0^2
\cs  + 2 a_0c_0 \cv, \qquad
\cv'=c_0d_0\crh + a_0b_0\cs +(a_0d_0+b_0c_0) \cv\, ,
\ee
and define
\be \label{esx4}
Q' = d_0 Q - b_0 P, \qquad P' = - c_0 Q + a_0 P\, .
\ee
Under this change of variables
\be \label{ejacob}
d\crh\wedge d\cs\wedge d\cv = d\crh'\wedge d\cs'\wedge d\cv'\, ,
\ee
\be \label{esx5}
Q_1\cdot P_2 - Q_2 \cdot P_1 = Q'\cdot P'\, ,
\ee
\be \label{esx5.5}
(Q_1,P_1) = (a_0Q', c_0Q'), \qquad (Q_2,P_2) = (b_0 P', d_0P')\, ,
\ee
and 
\be \label{esx6}
{1\over 2} \crh P^2 + {1\over 2} \cs Q^2 + \cv Q\cdot P
= {1\over 2} \crh' P^{\prime 2} + 
{1\over 2} \cs' Q^{\prime 2} + \cv' Q'\cdot P'\, .
\ee
Thus the jump in the index given in \refb{ex1} can be expressed as
\be \label{esx7}
(-1)^{Q'\cdot P'+1} \, Q'\cdot P'\, 
d_h(a_0Q', c_0Q') d_h(b_0P',d_0P')
\, .
\ee
Furthermore in these variables the pole at \refb{epole1} is at
$\cv'=0$. Thus we can identify \refb{esx7} 
with the residue of the
integrand from $\cv'=0$. Using \refb{ejacob}, \refb{esx6} the latter
may be expressed as
\be \label{esplit}
(-1)^{Q\cdot P+1} \int d\crh' d\cs' d \cv' e^{i\pi
(\crh' P^{\prime 2} + 
\cs' Q^{\prime 2} + 2\cv' Q'\cdot P') } \, {1\over \cp(\crh,\cs,\cv)}
\, ,
\ee
where the integration contour is around $\cv'=0$. We now note that
this result can be reproduced if we assume that 
near the pole
\refb{epole1} the 
partition function behaves as\footnote{This formula
suffers from the same type of subtleties described below
eq.\refb{ex5} with $(Q,P)$ replaced by $(Q',P')$  and
$(\crh,\cs,\cv)$ replaced by $(\crh',\cs',\cv')$.}
\be \label{ebe1}
{\cp(\crh,\cs,\cv)^{-1}} 
\propto \,
\{ {\phi_e(\cs';a_0,c_0)^{-1}} {\phi_m(\crh';b_0,d_0)^{-1}} 
\cv^{\prime -2} + \OO(\cv^{\prime 0})\}\, ,
\ee
where $1/\phi_{e,m}(\tau;k,l)$  denote
the partition functions of half BPS dyons in the set $\BBB$
such that
\ben \label{ebb1}
d_h(a_0 Q', c_0 Q') &=& {1\over T}\,
\int_{iM-T/2}^{ i M+T/2} d\tau\, e^{-i\pi Q^{\prime 2}\tau} 
{1\over \phi_e(\tau;a_0,c_0)}, \nonumber \\
d_h(b_0 P', d_0 P') &=& {1\over T'}\,
\int_{iM-T'/2}^{i M+T'/2} d\tau\, e^{-i\pi P^{\prime 2}\tau} 
{1\over \phi_m(\tau;b_0,d_0)}\, .
\een
The integration over $\tau$ run parallel to the real axis over unit
period with the imaginary part fixed at some large positive value
$M$.
Substituting \refb{ebe1} into \refb{esplit} and
picking up the residue from the pole at $\cv'=0$ we get the change
in the index to be
\be \label{bb2}
(-1)^{Q'\cdot P'+1} \, Q'\cdot P' \, d_h(a_0Q', c_0Q') 
d_h(b_0P',d_0P')\, ,
\ee
in agreement with \refb{esx7}.

To summarize, \refb{epole1} gives us the locations of the
zeroes of $\cp$, whereas  
eq.\refb{ex3} and more generally
\refb{ebe1} give us information about the behaviour of
$\cp$ near this zero. 
We shall now show that these results suggest additional symmetries
of $\cp$ of the type described in \refb{espec}.
Typically 
in any theory the partition functions of half BPS
states have modular properties. Let us for definiteness
consider the decay $(Q,P)\to (Q,0)+(0,P)$.
In this case the functions
$\phi_m(\crh)$ and $\phi_e(\cs)$ transform as modular forms
of a subgroup of $SL(2,\ZZZ)$ 
since they arise from quantization of a fundamental
string or a dual magnetic string. These relations
take the form
\be \label{ey1}
\phi_m((\alpha \crh + \beta) (\gamma\crh +\delta)^{-1})
= (\gamma\crh +\delta)^{k+2} \phi_m(\crh), \qquad
\phi_e((p\cs + q)(r\cs + s)^{-1}) = (r\cs + s)^{k+2}
\phi_e(\cs)\, ,
\ee
where $k$ is an integer specific to the theory under study,
and $\pmatrix{\alpha & \beta\cr \gamma & \delta}$ and
$\pmatrix{p & q\cr r & s}$ belong to appropriate subgroups
of $SL(2,\ZZZ)$. Given that $\phi_m$ and $\phi_e$ have these
symmetries, we conclude from \refb{ex3} that near $\cv=0$,
$\cp$ also has some additional symmetries. 
Even though there is no guarantee that these will be symmetries
of the full quarter BPS 
partition function, one could hope that some part
of these do lift to symmetries of the partition function and hence
of $\cp$.
Those which do 
can be represented by symplectic transformations
of the type \refb{espec} with
\be \label{ez1}
\pmatrix{A & B\cr C & D} = \pmatrix{\alpha & 0 & \beta & 0\cr
0 & 1 & 0 & 0\cr \gamma & 0 & \delta & 0\cr
0 & 0 & 0 & 1} \quad \hbox{and} \quad 
\pmatrix{A & B\cr C & D} = \pmatrix{1 & 0 & 0 & 0\cr
0 & p & 0 & q\cr 0 & 0 & 1 & 0\cr
0 & r & 0 & s}\, .
\ee
The first transformation generates
\be \label{ez2}
\crh \to {\alpha \crh+\beta \over \gamma\crh +\delta}, \quad
\cs \to \cs- {\gamma\cv^2 \over \gamma\crh +\delta}, \quad \cv \to {\cv
\over \gamma\crh +\delta}\, ,
\ee
while the second transformation generates
\be \label{ez3}
\crh \to \crh - {r\cv^2 \over r\cs + s}, \quad 
\cs \to {p\cs + q\over r\cs + s}, \quad \cv\to {\cv\over
r\cs + s}\, .
\ee
Both transformations leave the $\cv=0$ surface invariant. 
Furthermore
applying these transformations on \refb{espec} and using
\refb{ex3} near $\cv=0$
we generate the
transformation laws \refb{ey1}.

The symplectic transformations given in \refb{ez1}, if present, 
give us
the additional symmetries required to have $\cp$ transform
as a modular form under a non-trivial  subgroup
of $Sp(2,\ZZZ)$. We can use this to determine the subgroup
of $Sp(2,\ZZZ)$ under which we expect $\cp$ to transform as a
modular form and also the weight $k$ of the modular form. 
However since we do not know {\it a priori} which part of the
symmetry groups of $\phi_e$ and $\phi_m$ will lift to the symmetries
of $\cp$, this is not a fool proof method. Nevertheless these can
serve as guidelines for making an educated guess.

The behaviour of $\cp$ near the other zeroes given in \refb{epole1}
could provide us with additional information. If the zero of
$\cp$ at
\refb{epole1} is related to the one at $\cv=0$ by an S-duality
transformation then this information is not new. 
Since S-duality transformation acts by multiplying the 
matrix $\pmatrix{a_0 & b_0\cr c_0 & d_0}$ associated with a
wall from the left\cite{0702141}, this 
means that if $\pmatrix{a_0 & b_0\cr c_0
& d_0}$ itself is an S-duality transformation  then we do not
get a new information. To this we must also add the information
that multiplying
$\pmatrix{a_0 & b_0\cr c_0
& d_0}$ from
the right
by $\pmatrix{\lambda & 0 \cr 0 & \lambda^{-1}}$
for any $\lambda$ or by $\pmatrix{0 & 1\cr -1 & 0}$  
does not change the wall\cite{0702141}. However in many cases
even after imposing these equivalence relations one finds
inequivalent walls.\footnote{For example in $\ZZZ_6$ CHL
model with S-duality group $\Gamma_1(6)$
the wall corresponding to the matrix 
$\pmatrix{1 & 1\cr 2 & 3}$ is not equivalent to the wall
corresponding to $\pmatrix{1 & 0\cr 0 & 1}$. 
We shall discuss this example in some detail in 
\S\ref{sapp}.\label{ff1}} 
In such cases
the associated zero of $\cp$
cannot be related to the zero at $\cv=0$ by an S-duality
transformation, and we get new 
information.\footnote{Typically the
number of such additional zeroes is a finite number, providing us with
a finite set of additional information.}
Let $\pmatrix{a_0 & b_0\cr c_0 & d_0}$ be the matrix associated
with such a decay. If the corresponding partition functions
$\phi_m(\tau;b_0, d_0)$ and $\phi_e(\tau; a_0, c_0)$ 
have modular groups containing matrices
of the form $\pmatrix{\alpha_1 & \beta_1\cr \gamma_1 & \delta_1}$
and $\pmatrix{p_1 & q_1 \cr r_1 & s_1}$ respectively, then they may
be regarded as symplectic transformations generated by the matrices
\be \label{egensp}
\pmatrix{d_0 & b_0 &0&0 \cr c_0 & a_0 &0&0\cr
0&0& a_0 & -c_0\cr 0&0& -b_0 & d_0}^{-1}
\pmatrix{\alpha_1 & 0 & \beta_1 & 0\cr
0 & 1 & 0 & 0\cr \gamma_1 & 0 & \delta_1 & 0\cr
0 & 0 & 0 & 1}
\pmatrix{d_0 & b_0 &0&0 \cr c_0 & a_0 &0&0\cr
0&0& a_0 & -c_0\cr 0&0& -b_0 & d_0} 
\ee
and
\be \label{esecond}
\pmatrix{d_0 & b_0 &0&0 \cr c_0 & a_0 &0&0\cr
0&0& a_0 & -c_0\cr 0&0& -b_0 & d_0}^{-1}
\pmatrix{1 & 0 & 0 & 0\cr
0 & p_1 & 0 & q_1\cr 0 & 0 & 1 & 0\cr
0 & r_1 & 0 & s_1}
\pmatrix{d_0 & b_0 &0&0 \cr c_0 & a_0 &0&0\cr
0&0& a_0 & -c_0\cr 0&0& -b_0 & d_0}
\ee
respectively, acting on the original variables $(\crh,\cs,\cv)$.
Again we could hope that a part of this symmetry is a symmetry of
$\cp$.

We shall illustrate these by several examples in \S\ref{s3}.

\sectiono{Black hole entropy} \label{sblack}

Another set of constraints may be derived by requiring that the
formula for the index of quarter BPS states match the entropy
of the black hole carrying the same charges in the limit when the
charges are large. The consequences of this constraint have been
analyzed in detail in the 
past\cite{9607026,0412287,0510147,0605210} and reviewed in
\cite{0708.1270}. Hence our discussion will be
limited to a review of the salient features.

In the approximation where we keep the supergravity part of
the action containing only the two derivative terms, 
the black hole entropy is given by
\be \label{eblack1}
\pi\sqrt{Q^2 P^2 - (Q\cdot P)^2}\, .
\ee
In all known cases this result is reproduced by the asymptotic
behaviour of \refb{e2} for large charges. Furthermore the
leading asymptotic behaviour comes from the residue of the
partition function at the pole 
at\cite{9607026,0412287,0510147,0605210, 0708.1270}
\be \label{eblack2}
\crh \cs - \cv^2 + \cv=0\, ,
\ee
up to translations of $\crh$, $\cs$ and $\cv$ by their periods.
We shall assume that this result continues to hold in the general
case. Thus $\cp(\crh,\cs,\cv)$ must have a zero at \refb{eblack2}.
In order to find the behaviour of $\cp$ near this zero one needs to
know the first non-leading correction to the leading formula
\refb{eblack1} for the black hole entropy. 
{\it A priori} these corrections depend 
on the complete set of
four derivative terms in the quantum effective
action of the theory and are difficult to calculate.
However in all known examples one finds that the entropy
calculated just by including the Gauss-Bonnet term in the
effective action reproduces correctly the first non-leading
correction to the statistical entropy. If we assume that this
result continues to hold for a general theory then we can
use this to determine the behaviour of $\cp$ near 
\refb{eblack2} in terms of
the coefficient of the Gauss-Bonnet term in the effective action.

Since this procedure has been extensively studied in
\cite{9607026,0412287,0510147,0605210} and 
reviewed in \cite{0708.1270}, we shall only quote the result.
Typically the Gauss Bonnet term in the Lagrangian has the
form
\be \label{eblack3}
 \int d^4 x\, \sqrt{-\det g} \,
 \phi(a,S)\, 
\left\{ R_{\mu\nu\rho\sigma} R^{\mu\nu\rho\sigma}
- 4 R_{\mu\nu} R^{\mu\nu}
+ R^2
\right\} \, ,
\ee
where $\tau=a+iS$ is the axion-dilaton modulus and the function
$\phi(a,S)$ has the form
\be \label{eblack4}
\phi(a,S) = - {1\over 64\pi^2} \, \left( (k+2) \ln S 
+ \ln g(a+iS) + \ln g(-a+iS)\right)
+\hbox{constant}\, .
\ee
Here $k$ is the same integer that appeared
in \refb{espec} and $g(\tau)$ transforms as a modular
form of weight $k+2$ under the S-duality group. In a given theory
$g(\tau)$ can be calculated in string perturbation 
theory\cite{9610237,9708062}.
To the first non-leading order in the inverse power of charges, 
the effect of this term is to change the black hole entropy 
to\cite{0708.1270}
\be \label{eblack5}
S_{BH} = \pi \, \sqrt{Q^2 P^2 - (Q\cdot P)^2}
+64\, \pi^2\, \phi\left({Q\cdot P\over P^2},
{\sqrt{Q^2 P^2 - (Q\cdot P)^2}\over P^2}  \right) 
+\cdots\, 
\ee
The analysis of 
\cite{9607026,0412287,0510147,0605210, 0708.1270} 
shows that this behaviour can be reproduced
if we assume that near the zero at \refb{eblack2}
\be \label{eblack6}
\cp(\crh, \cs,\cv) \propto (2v-\rho-\sigma)^k\,
\{ v^2 \, g(\rho)\, g(\sigma) +  \OO(v^4)\}\, ,
\ee
where
\be \label{eblack7}
\rho = {\crh \cs - \cv^2\over \cs}, \quad
\sigma = {\crh \cs - (\cv-1)^2\over \cs}, \quad
v= {\crh \cs - \cv^2+\cv\over \cs}\, .
\ee
If we assume that eq.\refb{eblack6} holds in general, then it
gives us information about the behaviour of $\cp(\crh,\cs,\cv)$
near the zero at \refb{eblack2}. On the other hand if we can determine
$\cp$ from other considerations then the validity of \refb{eblack6}
would provide further evidence for the postulate that in $\NN=4$
supersymmetric string theories the Gauss-Bonnet term gives the
complete correction to black hole entropy to first non-leading order.

\sectiono{Examples} \label{s3}

\renewcommand{\theequation}
{\thesubsection.\arabic{equation}}

In this section we shall describe several applications of the
general procedure described in \S\ref{s2}. Some of them will
involve known cases and will provide a test for our procedure,
while others will be new examples where we shall 
derive a set of constraints on 
certain dyon partition functions which 
have not yet been computed from first principles.

\subsectiono{Dyons with unit torsion
in heterotic string theory on $T^6$} \label{s3.1}

We consider a dyon of charge $(Q,P)$ in the heterotic string theory
on $T^6$. $Q$ and $P$ take values in the 
Narain lattice $\Lambda$\cite{narain,nsw}.
Let $S^1$ and $\wt S^1$ be two circles of $T^6$, each labelled
by a coordinate with period $2\pi$  and let us denote
by $n',\wt n$ the momenta along $S^1$ and $\wt S^1$,
by $-w',-\wt w$ the  fundamental string winding numbers along
$S^1$ and $\wt S^1$, by $N',\wt N$ the Kaluza-Klein monopole
charges associated with $S^1$ and $\wt S^1$, and by $-W',-\wt W$
the H-monopole charges associated with $S^1$ and 
$\wt S^1$\cite{0708.1270}. Then
in the four dimensional subspace consisting of charge vectors
\be \label{ess1}
Q = \pmatrix{\wt n\cr n'\cr \wt w\cr w'}, \qquad
P = \pmatrix{\wt W\cr W'\cr \wt N\cr N'}\, ,
\ee
the metric $L$ takes the form 
\be \label{ess2}
L = \pmatrix{ 0 & I_2\cr I_2 & 0}\, ,
\ee
where $I_2$ denotes $2\times 2$ identity matrix. In this subspace
we consider a three parameter family of charge vectors $(Q,P)$ with
\be \label{ess3}
Q=\pmatrix{0 \cr m\cr 0 \cr -1}\, , \qquad P = 
\pmatrix{K \cr J \cr 1\cr 0}\, , \quad m, K, J\in\ZZZ\, .
\ee
This has
\be \label{ess4}
Q^2 = - 2 m, \quad P^2 = 2 K, \quad Q\cdot P = -J\, .
\ee
We shall identify this set of charge vectors as the set
$\AAA$.
As required, $Q^2$, $P^2$ and $Q\cdot P$ are independent linear
functions of $m$, $K$ and $J$ so that
for a pair of distinct values of $(m,K,J)$ we get a
pair of distinct values of $(Q^2, P^2, Q\cdot P)$.
All the charge vectors in this family
have unit torsion, \i.e.\ if we express the charges as linear combinations
$\sum Q_i e_i$ and $\sum P_i e_i$ of primitive basis elements
$e_i$ of the lattice $\Lambda$, then the torsion
\be \label{et1}
r(Q,P) \equiv \hbox{gcd}\{Q_i P_j - Q_j P_i\}\, ,
\ee
is equal to 1. In this case it is known that $Q^2$, 
$P^2$ and $Q\cdot P$
are the complete set of T-duality invariants\cite{0712.0043},
\i.e. beginning with a pair $(Q,P)$ with unit torsion we can reach
any other pair with unit torsion and same values of $Q^2$, $P^2$
and $Q\cdot P$ via a T-duality transformation. Since the set $\AAA$
contains all integer triplets $(Q^2/2, P^2/2, Q\cdot P)$ we conclude
that
the set $\BBB$ is the set of all $(Q,P)$ with unit torsion.
The corresponding partition function is 
known\cite{9607026} -- it is the inverse of the
weight ten Igusa cusp
form $\Phi_{10}$ of the full $Sp(2,\ZZZ)$ group.

We shall now examine how $\Phi_{10}$ satisfies the various
constraints derived in the previous sections. First of all
note that
since S-duality transformation does not change the torsion $r$, the
full $SL(2,\ZZZ)$ group is a symmetry of this set. Furthermore
in this set $Q^2/2$, $P^2/2$ and $Q\cdot P$ are all quantized in
integer units. Thus the partition function is invariant under
translation of $\crh$, $\cs$ and $\cv$ by arbitrary integer units.
These correspond to symplectic transformations of the form
\ben \label{ecorr}
&& \pmatrix{d & b &0&0 \cr c & a &0&0\cr
0&0& a & -c\cr 0&0& -b & d}, \quad \hbox{and} \quad
\pmatrix{1 & 0 & a_1 & a_3\cr
0 & 1 & a_3 & a_2\cr 0 & 0 & 1 & 0\cr 0 & 0 & 0 & 1}
\nonumber \\ &&
\pmatrix{a & b\cr c & d}\in SL(2,\ZZZ), \quad a_1,a_2,a_3\in\ZZZ\, .
\een
Clearly each of these transformations belong to
$Sp(2,\ZZZ)$ and is a symmetry of $\Phi_{10}$.

Next we turn to the constraints from the
wall crossing formula. 
In this case all the walls are related by S-duality transformation to
the wall corresponding to the decay $(Q,P)\to (Q,0)+(0,P)$.
So it is sufficient to study the consequences of
the wall crossing formula at this wall. 
Clearly $Q^2$ and $P^2$ given in \refb{ess4} are uncorrelated.
Furthermore in heterotic string theory on $T^6$ all $Q$'s with
a given $Q^2$ are related by $T$-duality transformation\cite{wall}.
The same is true for $P$. Thus the subtleties mentioned below
eq.\refb{ex5} are absent, and
the behaviour of 
$\cp(\crh,\cs,\cv)$ near $\cv=0$ is expected to be
given by \refb{ex3}. In this case 
both the electric and the magnetic half-BPS partition functions
are given by $\eta(\tau)^{-24}$
where $\eta$ denotes the Dedekind function. Thus we have,
as a consequence of the wall crossing formula,
\be \label{et2a}
\cp(\crh,\cs,\cv) \propto \{ \cv^2 \, (\eta(\crh))^{24}\,  (\eta(\cs))^{24}
+\OO(\cv^4)\}\,.
\ee
$\eta(\tau)^{24}$ transforms as a modular form of weight 12 
under an $SL(2,\ZZZ)$ transformation. {}From eqs.\refb{ez1} 
it follows that these $SL(2,\ZZZ)$ transformations
may be regarded as the
following symplectic transformations of $\crh$, $\cs$, $\cv$
\ben \label{ecorr2}
&&\pmatrix{\alpha & 0 & \beta & 0\cr
0 & 1 & 0 & 0\cr \gamma & 0 & \delta & 0\cr
0 & 0 & 0 & 1} \quad \hbox{and} \quad
\pmatrix{1 & 0 & 0 & 0\cr
0 & p & 0 & q\cr 0 & 0 & 1 & 0\cr
0 & r & 0 & s}\nonumber \\ &&
\pmatrix{\alpha & \beta\cr \gamma & \delta}\in SL(2,\ZZZ),
\quad \pmatrix{p & q\cr r & s}\in SL(2,\ZZZ)\, .
\een
Furthermore $\cp$ should have weight $12-2=10$.

Let us now compare these with the known 
properties of $\Phi_{10}$. $\Phi_{10}(\crh,\cs,\cv)$ is indeed
known to have the factorization property \refb{et2a}.
Furthermore since $\Phi_{10}$ transforms as a modular form of weight
10 under the full $Sp(2,\ZZZ)$ group, and since \refb{ecorr2}
are $Sp(2,\ZZZ)$ matrices, they represent symmetries of
$\Phi_{10}$. Thus we see that in this case the full set of
symmetries of $\phi_m$ and $\phi_e$ lift to symmetries of
$\cp$. 
It is worth noting that the matrices given in
\refb{ecorr} and \refb{ecorr2} generate the full $Sp(2,\ZZZ)$ group.
Thus in this case by assuming that the full modular 
groups of $\phi_e$ and
$\phi_m$ lift to symmetries of the partition function we could
determine the symmetries of the partition function.

Finally let us consider the constraints coming from the 
knowledge of black hole entropy. In this 
case the function $g(\tau)$ appearing in \refb{eblack4} is given by
$\eta(\tau)^{24}$.  Thus \refb{eblack6} takes the
fom
\be \label{eblack6a}
\cp(\crh, \cs,\cv) \propto (2v-\rho-\sigma)^{10}\,
\{ v^2 \, \eta(\rho)^{24}\, 
\eta(\sigma)^{24} +  \OO(v^4)\}\, ,
\ee
where $(\crh,\cs,\cv)$ and $(\rho,\sigma,v)$ are related via
eq.\refb{eblack7}.
The Siegel modular form $\Phi_{10}(\crh,\cs,\cv)$
satisfies these properties. In fact since \refb{eblack7}
represents an $Sp(2,\ZZZ)$ transformation, the property
\refb{eblack6a} of $\cp(\crh, \cs,\cv)$ follows from the
factorization property \refb{et2a}. This however will not be
the case in a more generic situation.

\subsectiono{Dyons with unit torsion and even $Q^2/2$
in heterotic  on $T^6$} \label{s3.2}

We now consider again heterotic string theory on $T^6$, but
choose the set $\AAA$ to be collection of $(Q,P)$ of the form:
\be \label{ess3a}
Q=\pmatrix{0 \cr 2m\cr 0 \cr -1}\, , \qquad P = 
\pmatrix{K \cr J \cr 1\cr 0}\, , \quad m, K, J\in\ZZZ\, .
\ee
This has
\be \label{ess4a}
Q^2 = - 4 m, \quad P^2 = 2 K, \quad Q\cdot P = -J\, .
\ee
We note that all the
charge vectors have $Q^2/2$ even. Since $Q^2$
is T-duality invariant, any other charge vector which can be
obtained from this one by a T-duality transformation has
$Q^2/2$ even. Thus the set $\BBB$ now consists of charge vectors
which have even $Q^2/2$ and arbitrary integer values of
$P^2/2$ and $Q\cdot P$.
Since this set $\BBB$ is a subset of charges for which the
spectrum was analyzed in \S\ref{s3.1} we do not expect to derive
any new results. Nevertheless we have chosen this example 
as this will serve as a useful guide to our analysis in later
sections.

We first note that
the quantization conditions of $Q^2$, $P^2$
and $Q\cdot P$ imply the following periods of the partition
function:
\be \label{ess6}
(\crh,\cs,\cv)\to (\crh+a_1, \cs+a_2, \cv+a_3), \quad
a_1 \in \ZZZ, \quad a_2 \in {1\over 2}\ZZZ, \quad a_3
\in \ZZZ\, .
\ee
The period along $\cs$ is not an integer. We can
remedy this by
defining 
\be \label{ess7}
Q_s = Q/2\, , \qquad \cs_s = 4\cs, \qquad \cv_s = 2\cv\, .
\ee
so that $Q_s^2/2$ and $Q_s\cdot P$
are now quantized in half integer units. The periods 
$(\wt a_1,\wt a_2,\wt a_3)$ of the variables
$(\crh,\cs_s,\cv_s)$ conjugate to $(P_s^2/2,Q_s^2/2,Q_s\cdot P_s)$
are now integers, given by,
\be \label{ess7a}
\wt a_1 \in \ZZZ, \quad \wt a_2 \in 2\ZZZ, \quad \wt a_3 \in 2
\ZZZ\, .
\ee
The dyon partition function in this case can be easily calculated
from the one for \S\ref{s3.1} by taking into account the
evenness of $Q^2/2$. This
amounts
to adding to the original partition function another term where
$\cs$ is shifted by $1/2$. Thus we have
\be \label{eus1}
{1\over \cp(\crh,\cs,\cv)} = {1\over 2}
\, \left( {1\over \Phi_{10}(\crh, \cs, 
\cv)}
+ {1\over \Phi_{10}(\crh, \cs+{1\over 2},
\cv)}\right)\, ,
\ee
or, in terms of the rescaled variables,
\be \label{est5}
{1\over \cp(\crh,\cs,\cv)} ={1\over 2}
\left( {1\over \Phi_{10}(\crh, {1\over 4}\cs_s, {1\over 2}\cv_s)}
+ {1\over \Phi_{10}(\crh, {1\over 4}\cs_s+{1\over 2}, 
{1\over 2}\cv_s)}\right)\, .
\ee
Let us determine the symmetries of this partition function. For
this it will be useful to work in terms of the original
unscaled variables
$(\crh, \cs,\cv)$ and at the
end go back to the rescaled variables. 
The first term on the right hand side of \refb{eus1}
has the usual $Sp(2,\ZZZ)$ symmetries acting
on the variables $(\crh, \cs, 
\cv)$. However not all of these are symmetries of the second
term. Given an $Sp(2,\ZZZ)$ matrix
$\pmatrix{a_1 & b_1 & c_1 & d_1\cr
a_2 & b_2 & c_2 & d_2\cr
a_3 & b_3 & c_3 & d_3\cr
a_4 & b_4 & c_4 & d_4}$, it is a symmetry of the
second term provided its action on 
$(\crh, \cs, 
\cv)$ can be regarded as an $Sp(2,\ZZZ)$ action 
$\pmatrix{a_1' & b_1' & c_1' & d_1'\cr
a_2' & b_2' & c_2' & d_2'\cr
a_3' & b_3' & c_3' & d_3'\cr
a_4' & b_4' & c_4' & d_4'}$
on
$(\crh, \cs+{1\over 2}, 
\cv)$ followed by a translation on $\cs$ by
1/2. Since a translation of $\cs$ by 1/2 can be
regarded as a symplectic transformation with the matrix
$\pmatrix{1 & 0 & 0 & 0\cr 0 & 1 & 0 & 1/2\cr 0 & 0 & 1 & 0\cr
0 & 0 & 0 & 1}$, the above condition takes the form:
\be \label{eresid}
\pmatrix{1 & 0 & 0 & 0\cr 0 & 1 & 0 & 1/2\cr 0 & 0 & 1 & 0\cr
0 & 0 & 0 & 1}
\pmatrix{a_1 & b_1 & c_1 & d_1\cr
a_2 & b_2 & c_2 & d_2\cr
a_3 & b_3 & c_3 & d_3\cr
a_4 & b_4 & c_4 & d_4}
= \pmatrix{a_1' & b_1' & c_1' & d_1'\cr
a_2' & b_2' & c_2' & d_2'\cr
a_3' & b_3' & c_3' & d_3'\cr
a_4' & b_4' & c_4' & d_4'} 
\pmatrix{1 & 0 & 0 & 0\cr 0 & 1 & 0 & 1/2\cr 
0 & 0 & 1 & 0\cr
0 & 0 & 0 & 1}\, .
\ee
This gives
\be \label{eresid2}
\pmatrix{a_1' & b_1' 
& c_1' & d_1'\cr
a_2' & b_2' & c_2' & d_2'\cr
a_3' & b_3' & c_3' & d_3'\cr
a_4' & b_4' & c_4' & d_4'} 
=\pmatrix{a_1 & b_1 & c_1 & d_1 - {1\over 2}b_1\cr
a_2+{1\over 2} a_4 & b_2+{1\over 2} b_4 
& c_2+{1\over 2} c_4 & d_2+{1\over 2} (d_4-b_2) -{1\over 4} b_4\cr
a_3 & b_3 & c_3 & d_3-{1\over 2} b_3\cr
a_4 & b_4 & c_4 & d_4-{1\over 2} b_4}\, .
\ee
The coefficients $a_i$, $b_i$, $c_i$ and $d_i$ are integers.
Requiring that there exist integer $a_i'$, $b_i'$, $c_i'$ and
$d_i'$ satisfying the above constraints we get further conditions
on $a_i$, $b_i$, $c_i$ and $d_i$. These take the following form:
\be \label{eresid3}
a_4, b_4, c_4, b_1,
b_3\in 2\ZZZ, \quad  b_4 - 2(d_4 - b_2) \in 4\ZZZ\, .
\ee
On the other hand the requirement that the original matrix
$\pmatrix{a_1 & b_1 & c_1 & d_1\cr
a_2 & b_2 & c_2 & d_2\cr
a_3 & b_3 & c_3 & d_3\cr
a_4 & b_4 & c_4 & d_4}$ is symplectic, together with the
first set of conditions given in \refb{eresid3}, can be used to 
show that $b_2$ and $d_4$
are both odd. As a result $(b_2-d_4)$ is even, and hence
$b_4$ must be a multiple of 4 in order to satisfy \refb{eresid3}.
Thus we have
\be \label{eresid4}
a_4 = 2 \wh a_4, \quad b_4 = 4\wh b_4, \quad
c_4 = 2 \wh c_4, \quad b_1 = 2 \wh b_1, \quad
b_3 = 2\wh b_3, \quad \wh a_4, \wh b_4, \wh c_4, \wh b_1,
\wh b_3\in \ZZZ\, .
\ee
This determines the subgroup of $Sp(2,\ZZZ)$ which leaves the
individual terms in \refb{eus1} invariant. To this we must add the
additional element corresponding to 
$\cs \to \cs+{1\over 2}$ which exchanges the two terms
in \refb{eus1}. This corresponds to the symplectic transformation
\be \label{eresid5}
\pmatrix{1 & 0 & 0 & 0\cr 0 & 1 & 0 & 1/2\cr 0 & 0 & 1 & 0\cr
0 & 0 & 0 & 1}\, .
\ee
The full symmetry group is then generated by the matrices:
\be \label{eresid6}
\pmatrix{a_1 & 2\wh b_1 & c_1 & d_1\cr
a_2 & b_2 & c_2 & d_2\cr
a_3 & 2\wh b_3 & c_3 & d_3\cr
2\wh a_4 & 4\wh b_4 & 2\wh c_4 & d_4} \quad \hbox{and} \quad
\pmatrix{1 & 0 & 0 & 0\cr 0 & 1 & 0 & 1/2\cr 0 & 0 & 1 & 0\cr
0 & 0 & 0 & 1}\, .
\ee
We can easily determine how these transformations
act on the rescaled variables $(\crh,\cs_s,\cv_s)$.
This is done with the help of conjugation by the
symplectic matrix
\be \label{econj1}
\pmatrix{1 & 0 & 0 & 0\cr 0 & 2 & 0 & 0\cr 0 & 0 & 1 & 0\cr
0 & 0 & 0 & 1/2}\, 
\ee
relating $(\crh,\cs,\cv)$ to $(\crh,\cs_s,\cv_s)$.
This converts the generators given in \refb{eresid6} to
\be \label{econj2}
\pmatrix{a_1 & \wh b_1 & c_1 & 2d_1\cr
2a_2 & b_2 & 2c_2 & 4d_2\cr
a_3 & \wh b_3 & c_3 & 2d_3\cr
\wh a_4 & \wh b_4 & \wh c_4 & d_4} \quad \hbox{and} \quad
\pmatrix{1 & 0 & 0 & 0\cr 0 & 1 & 0 & 2\cr 0 & 0 & 1 & 0\cr
0 & 0 & 0 & 1}\, .
\ee
We now note that all the matrices appearing in \refb{econj2}
have the form
\be \label{econj3}
\pmatrix{* & * & * & 0\cr 0 & * & 0 & 0\cr * & * & * & 0\cr
* & * & * & *} \quad \hbox{mod 2}\, ,
\ee
with $*$ denoting an arbitrary integer subject to the condition
that \refb{econj3} describes a symplectic matrix. Furthetmore the
set of matrices \refb{econj3} are closed under matrix multiplication.
Thus the
group generated by the matrices \refb{econj2} is
contained in the group $\check G$ consisting of $Sp(2,\ZZZ)$
matrices of the form \refb{econj3}. It is in fact easy to show that
the group generated by the matrices \refb{econj2} is the whole
of $\check G$, \i.e.\ any element of $\check G$ given in
\refb{econj3} can be written as a product
of the elements given in \refb{econj2}.

We shall now set aside this result for a while and study the implications
of S-duality symmetry and the wall crossing formula on the partition
function. The eventual goal is to test the conclusions drawn from the
general arguments along the lines of  \S\ref{sdual} and \S\ref{swall}
against the known results for $\cp$ given above. 
It follows from \refb{e3.5} and \refb{ess4a}
that in order that an S-duality 
transformation generated by $\pmatrix{a & b\cr c & d}$ takes an
arbitrary element of the set $\BB$ to another element of the
set $\BB$ we must have $b$ even. Thus S-duality transformations
which preserve the set $\BB$ take the form:
\be \label{ess5}
Q\to Q'' = a Q + b P, \quad P\to P''=cQ + d P, \quad
a,c,d\in\ZZZ, \quad b\in 2\ZZZ, \quad ad-bc=1\, .
\ee
On the original variables $(\crh,\cs,\cv)$ the associated transformation
can be represented by the symplectic matrix \refb{e13}. After
conjugation by the matrix \refb{econj1} 
we get the symplectic
matrix acting on the rescaled variables $(\crh,\cs_s,\cv_s)$:
\be \label{esympaa}
\pmatrix{d & \wt b &0&0 \cr \wt c & a &0&0\cr
0&0& a & -\wt c\cr 0&0& -\wt b & d}\,, \quad a,\wt b\equiv b/2,
d\in\ZZZ, \quad \wt c\equiv 2c\in 
2\ZZZ\, .
\ee
This clearly has the form given in \refb{econj3}. Also the
periodicities along the $(\crh,\cs_s,\cv_s)$ directions, as given in
\refb{ess7a}, are represented by the symplectic transformation
\be \label{esya1}
\pmatrix{1 & 0 & \wt a_1 & \wt a_3\cr 0 & 1 & \wt
a_3 & \wt a_2\cr
0 & 0 & 1 & 0\cr 0 & 0 & 0 & 1}, \qquad \wt a_1\in\ZZZ, 
\quad \wt a_2, \wt a_3 \in 2\ZZZ\, .
\ee
These also are of the form given in \refb{econj3}.

Next we turn to the information obtained from the wall
crossing relations.  Consider first the wall associated with 
decay $(Q,P)\to (Q,0) + (0, P)$, -- this controls the
behaviour of $\cp$ near $\cv=0$ via eq.\refb{ex3}. 
Since $Q^2=-4m$ and $P^2=2K$ can vary 
independently inside
the set $\AAA$, and since any
two charge vectors of the same norm
can be related by a T-duality transformation\cite{wall},
there is no subtlety of the type described below
\refb{ex5}.
The inverse of the magnetic
partition function $\phi_m$ entering \refb{ex3} is the same as
the one that appeared in \refb{et2a}:
\be \label{est1}
\phi_m(\crh) = (\eta(\crh))^{24}\, .
\ee
The electric partition function gets modified from the
corresponding expression given in \refb{et2a}
due to 
the fact that we are only including even
$Q^2/2$ states. As a result the partition function now
becomes ${1\over 2}
\left(\eta(\cs)^{-24} + 
\eta\left(\cs+{1\over 2}\right)^{-24}\right)$.
Replacing $\cs$ by $\cs_s/4$
we get
\be \label{est2}
\phi_e(\cs)^{-1} = {1\over 2}\left(\eta\left({\cs_s\over 4}
\right)\right)^{-24} + {1\over 2}
\left(\eta\left({\cs_s\over 4}+{1\over 2}\right)\right)^{-24}\, .
\ee
This leads to the following behaviour of $\cp$ near
$\cv_s=0$:
\be \label{efactor1}
\cp(\crh,\cs,\cv) \propto \left[
\cv_s^2 \, \eta(\crh)^{24} \, \left\{
\left(\eta\left({\cs_s\over 4}
\right)\right)^{-24} + 
\left(\eta\left({\cs_s\over 4}+{1\over 2}\right)\right)^{-24}
\right\}^{-1} + \OO(\cv_s^4)
\right]
\ee
$\cp$ given in \refb{est5} can be shown to satisfy this property.

$\phi_m(\crh)$ given in 
\refb{est1} transforms  as a modular form of weight 12 under
\be \label{est1.5}
\crh\to {\alpha \crh +\beta\over \gamma\crh + \delta}, 
\qquad \pmatrix{\alpha &\beta \cr\gamma & \delta}
\in SL(2,\ZZZ)\, .
\ee
On the other hand
$\phi_e(\cs)$ given in \refb{est2} can be shown to
transform as a modular form of weight
12 under
\be \label{est3}
\cs_s \to {p\cs_s + q\over r\cs_s + s}, \quad \pmatrix{ p & q\cr r & s}
\in\Gamma^0(2)\, ,
\ee
\i.e. $SL(2,\ZZZ)$ matrices with $q$ even.
\refb{est1.5} and \refb{est3} can be represented as symplectic 
transformations of $(\crh,\cs_s,\cv_s)$ generated by the 
$Sp(2,\ZZZ)$ matrices
\be \label{es1a}
\pmatrix{\alpha & 0 & \beta & 0\cr
0 & 1 & 0 & 0\cr \gamma & 0 & \delta & 0\cr
0 & 0 & 0 & 1} \quad \hbox{and} \quad 
\pmatrix{1 & 0 & 0 & 0\cr
0 & p & 0 & q\cr 0 & 0 & 1 & 0\cr
0 & r & 0 & s}\, , \quad  q\in 2 \ZZZ, \quad
\alpha,\beta,\gamma,\delta,p,r,s\in\ZZZ\, .
\ee
We now note that these transformations fall in the class given in
\refb{econj3}. Thus in this case
the modular symmetries of the half-BPS
partition function associated with pole at $\cv=0$ are lifted to
symmetries of the full partition function.

In this case there is one additional wall which is not related to the
wall considered above by the $\Gamma^0(2)$ S-duality 
transformation \refb{ess5} acting on the original variables.
This corresponds to the decay  $(Q,P)\to (Q-P, 0) +(P,P)$.
Comparing this with \refb{ede1} we see that here
\be \label{esv1}
\pmatrix{a_0 & b_0\cr c_0 & d_0} = \pmatrix{1 & 1\cr 
0 & 1}\, .
\ee
Following \refb{esx3}, \refb{esx4} and the relationship
\refb{ess7} between the original variables and the rescaled variables
we have
\be \label{esv2}
\crh' = \crh + {1\over 4}\cs_s +\cv_s, \quad \cs' = {1\over 4}{\cs_s}, 
\quad
\cv' = {1\over 2} \cv_s +{1\over 4} \cs_s\, ,
\ee
\be \label{esv3}
Q' = Q-P, \quad P' = P\, .
\ee
Thus the pole 
of the partition function is at $\cv_s + {1\over 2}\cs_s=0$.
Furthermore 
since from the relations \refb{ess4a} we see that the allowed
values of $(Q-P)^2/2=J+K-2m$ and 
$P^2/2=K$ are uncorrelated and can take
arbitrary integer values,  
it follows from \refb{ebe1} that
at this zero $\cp$ goes as
\be \label{esv4}
\cp(\crh,\cs_s,\cv_s) \propto (2\cv_s +\cs_s)^2 \, \phi_m\left(\crh
+{1\over 4}\cs_s+\cv_s;{1}, 1\right) \, 
\phi_e\left({1\over 4}\cs_s; 1,0\right) +
\OO\left((2\cv_s+\cs_s)^4\right)\, .
\ee
$\phi_m(\tau; {1}, 1)$ denotes the partition 
function of half-BPS
states carrying charges $(P,  P)$, 
with $\tau$ being conjugate to the variable 
$P^{\prime 2}/2=P^2/2$. Thus
we have $\phi_m(\tau; {1}, 1)=(\eta(\tau))^{24}$. On the
other hand $(\phi_e(\tau;1,0))^{-1}$ is the partition function of half
BPS states carrying charges $(Q',0)=(Q-P, 0)$ 
 with $\tau$ being conjugate to $Q^{\prime 2}/2
= (Q-P)^2/2$. Since $(Q-P)^2/2=(-2m+K+J)$ can take arbitrary
integer values, the corresponding 
partition function is also given by
$\eta(\tau)^{-24}$. 
Thus 
we have near $(\cs_s+2\cv_s)=0$
\be \label{efactor2}
\cp(\crh,\cs_s,\cv_s) \propto \left\{ (2\cv_s+\cs_s)^2 \eta\left(
\crh + {1\over 4} \cs_s +\cv_s\right)^{24} 
\eta\left({\cs_s\over 4}\right)^{24}
+ \OO((2\cv_s+\cs_s)^4)
\right\}
\ee
$\cp$ given in \refb{est5} can be shown to satisfy this property.

$\phi_m(\tau;{1}, 1)$ transforms as a modular
form of weight 12 under
$\tau\to (\alpha_1\tau +\beta_1)/(\gamma_1\tau+\delta_1)$ with
$\alpha_1,\beta_1,\gamma_1, \delta_1\in \ZZZ$, $\alpha_1\delta_1
-\beta_1\gamma_1=1$. On the other hand $\phi_e(\tau;1,0)$ transforms
as a modular form of weight 12 under 
$\tau\to (p_1\tau+q_1)/(r_1\tau+s_1)$
with $p_1,q_1,r_1,s_1\in\ZZZ$,  
$p_1s_1 -q_1r_1=1$. 
Using \refb{egensp}, \refb{esecond} and \refb{econj1}
we see that the the action of these transformations
on the variables $(\crh,\cs_s,\cv_s)$ may be represented by the
symplectic matrices
\ben \label{esv6}
&& \pmatrix{\alpha_1 & (\alpha_1-1)/2 & \beta_1 & 0\cr
0 & 1 & 0 & 0\cr \gamma_1 & \gamma_1/2 & \delta_1 & 0\cr
\gamma_1/2 & \gamma_1/4 & (\delta_1 -1)/2 & 1}, \quad
\pmatrix{1 & (1 - p_1)/2 & q_1  & -2q_1  \cr 
0 & p_1 & -2q_1  & 4q_1\cr  0 & 0 & 1 & 0\cr 0 & r_1/4 
& (1-s_1)/2 & s_1}\,  . \nonumber \\
\een
By comparing with the matrices given in \refb{econj3} we see
however that the transformations \refb{esv6} generates symmetries
of the full partition function only after we impose the additional
constraints
\be \label{esksk1}
r_1, \gamma_1\in 4\ZZZ\, .
\ee
Thus here we encounter a case where only a subset of 
the symmetries of the partition function near a pole is lifted to
a full symmetry of the partition function. By examining the details
carefully one discovers that in this case the 
pole comes from the first term 
in \refb{est5}. Whereas this term displays the full symmetry
given in \refb{esv6}, requiring that the other term also transforms
covariantly under this symmetry generates the additional
restrictions given in \refb{esksk1}.

Finally we turn to the constraint from black hole entropy. As in
\S\ref{s3.1}, in this case we have $g(\tau)=\eta(\tau)^{24}$
in \refb{eblack4}.  Thus \refb{eblack6} takes the
form
\be \label{eblack6b}
\cp(\crh, \cs,\cv) \propto (2v-\rho-\sigma)^{10}\,
\{ v^2 \, \eta(\rho)^{24}\, 
\eta(\sigma)^{24} +  \OO(v^4)\}\, ,
\ee
where  $(\crh,\cs,\cv)$ and $(\rho,\sigma,v)$ are related via
\refb{eblack7}.
$\cp(\crh,\cs,\cv)$ given in \refb{eus1} can be shown to satisfy this
property. In fact the relevent pole of $\cp^{-1}$ comes from the first
term on the right hand side of \refb{eus1}. The location of
the zeroes of $\Phi_{10}$ are given in \refb{eapl5a}, 
and it follows from this
that the second term does not have a pole at
$v=0$.

\subsectiono{Dyons of torsion 2 in heterotic string theory 
on $T^6$} 
\label{s3.4}

We consider again heterotic string theory on $T^6$ and take the
set $\AAA$ to consist of charge vectors of the form
\be \label{esa1}
Q=\pmatrix{1 \cr 2m+1\cr 1 \cr 1}\, , \qquad P = 
\pmatrix{2K+1 \cr 2J+1 \cr 1\cr -1}\, , \quad m, K, J\in\ZZZ\, .
\ee
This has
\be \label{esa2}
Q^2 = 4(m+1), \quad P^2 = 4(K-J), \quad Q\cdot P = 2(K+J-m+1)\, .
\ee
Furthermore gcd$\{Q_i P_j - Q_j P_i\}$=2. Thus we have a
family of charge vectors with torsion
2.  It was shown in \cite{0712.0043,0801.0149} that for $r=2$
there are three T-duality
orbits for given $(Q^2, P^2, Q\cdot P)$ -- in the first
$Q$ is twice a primitive lattice vector, in the second $P$ is
twice a primitive lattice vector and in the third both $Q$ and $P$
are primitive but $Q\pm P$ are twice primitive lattice vectors.
The dyon charges given in \refb{esa1} are clearly of the third
kind.
In the notation of \cite{0801.0149} the 
discrete T-duality invariants of
these charges are $(r_1=1, r_2=1, r_3=2, u_1=1)$.
Note that as we vary $m$, $J$ and $K$, $Q^2/2$ and $P^2/2$ take
all possible even values and $Q\cdot P$ takes all possible
values subject to the restriction that $Q\pm P$ are
twice primitive lattice vectors. The latter condition requires
$Q\cdot P$ to be even and
$Q\cdot P -{1\over 2}Q^2 - {1\over 2}P^2$ to be a multiple
of four. It now follows from the result of 
\cite{0712.0043,0801.0149} that the T-duality
orbit $\BBB$ of the set $\AAA$ consists of all the pairs
$(Q,P)$ with $(r_1=1, r_2=1, r_3=2, u_1=1)$ and
even values of $Q^2/2$, $P^2/2$.

Since $Q^2/2$, $P^2/2$ and $Q\cdot P$ are all
even and $Q^2 + P^2 + 2Q\cdot P$ is a multiple of 8,
it is natural to introduce  new charge vectors and variables
\be \label{escale2}
Q_s\equiv Q/2, \quad P_s \equiv P/2, \quad \crh_s\equiv 
4\crh, \quad
\cs_s \equiv 4\cs, \quad \cv_s\equiv 4\cv\, ,
\ee 
so that we have
\be \label{esa3}
{1\over 2}Q_s^2 = {1\over 2}
(m+1), \quad {1\over 2} P_s^2 = {1\over 2}
(K-J), \quad Q_s\cdot P_s 
= {1\over 2} (K+J-m+1)\, ,
\ee
quantized in half integer units subject to the constraint that
\be \label{econstr}
{1\over 2}Q_s^2 + {1\over 2} P_s^2 +  Q_s\cdot P_s
= K + 1\, ,
\ee
is an integer. 
Since $(\crh_s,\cs_s,\cv_s)$ are conjugate to
$(P_s^2/2, Q_s^2/2, Q_s\cdot P_s)$,
the partition function \refb{e1} will be periodic under
\be \label{eper}
(\crh_s, \cs_s, \cv_s) \to (\crh_s+2, \cs_s, \cv_s), \,
(\crh_s, \cs_s+2, \cv_s), \, (\crh_s, \cs_s, \cv_s+2), 
\, (\crh_s+1, \cs_s+1, \cv_s+1)\, .
\ee
The group generated by these transformations
can be collectively represented by symplectic matrices
of the form
\be \label{ecollect}
\pmatrix{1 & 0 & \wt a_1 & \wt a_3\cr
0 & 1 & \wt a_3 & \wt a_2\cr 0 & 0 & 1 & 0
\cr 0 & 0 & 0 & 1}, \quad
\wt a_1, \wt a_2, \wt a_3\in\ZZZ, \quad 
\wt a_1+\wt a_2, \wt a_2+\wt a_3, \wt a_1+\wt a_3\in 2\ZZZ\, ,
\ee
acting on the variables $(\crh_s,\cs_s,\cv_s)$.
For future reference we note that 
the change of variables from $(\crh,\cs,\cv)$ to $(\crh_s,\cs_s,\cv_s)$
can be regarded as a symplectic transformation of the form
\be \label{esym1}
\pmatrix{2 & 0 & 0 & 0 \cr  0 & 2 & 0 & 0\cr 0 & 0 & 1/2 & 0\cr
0 & 0 & 0 & 1/2}\, .
\ee

We now need to determine the subgroup of the S-duality
group that leaves the set $\BBB$ invariant. If we did not
have the restriction that $Q^2/2$ and $P^2/2$ are even, then this
subgroup would consist of $SL(2,\ZZZ)$ matrices of the
form $\pmatrix{a & b\cr c & d}$ subject to the restriction
$a+b\in 2\ZZZ+1$ and $c+d\in 2\ZZZ+1$\cite{0801.0149}, --
these conditions 
guarantee  that
the new charge vectors $(Q'',P'')$ are each primitive and
hence have the same set of discrete 
T-duality invariants 
$(r_1=1, r_2=1, r_3=2, u_1=1)$.
 We shall now
argue that the same subgroup also leaves the set $\BBB$ invariant.
For this we need to note that if we begin with a $(Q,P)$ for which
$Q^2/2$, $P^2/2$ and $Q\cdot P$ are all even then their
S-duality transforms given in
\refb{e3.5} will automatically have the same properties. 
Thus requiring the transformed
pair $(Q'',P'')$ to have even $Q^{\prime\prime2}/2$ and 
$P^{\prime\prime2}/2$, as is required for $(Q'', P'')$ to
belong to the set $\BBB$,
does not put any additional restriction on the S-duality
transformations. Since both $Q$ and $P$ are
scaled by the same amount to get the rescaled charges
$Q_s$ and $P_s$, the S-duality group action on $(Q_s, P_s)$
is identical to that on $(Q,P)$ and hence its action on 
$(\crh_s,\cs_s,\cv_s)$ is identical to that on $(\crh,\cs,\cv)$.
Using \refb{e13} we see that
the  representations of these symmetries as symplectic matrices
are given by
\be \label{ete2s}
\pmatrix{d & b &0&0 \cr c & a &0&0\cr
0&0& a & -c\cr 0&0& -b & d}, \qquad a,b,c,d\in\ZZZ, \quad ad-bc=1,
\quad a+c\in 2\ZZZ+1, \quad b+d\in 2\ZZZ+1\, ,
\ee
acting on the variables $(\crh,\cs,\cv)$ and also on
$(\crh_s,\cs_s,\cv_s)$.

Next we turn to the constraints from the wall crossing formula.
We begin with the wall associated with the decay $(Q,P)\to
(Q,0)+(0,P)$, --
this controls the behaviour of $\cp$ at $\cv=0$.
The analysis is straightforward. We note that 
both electric and magnetic partition functions involve
summing over all possible
even $Q^2/2$ and $P^2/2$ values.  An analysis
similar to the one leading to \refb{est2} give
\be \label{ex11}
\phi_e(\cs)^{-1}=
{1\over 2} \left\{   \eta\left({\cs_s\over 4}\right)^{-24}
+ \eta\left({\cs_s+2\over 4}\right)^{-24} 
\right\} \, ,
\ee
and
\be \label{ex12}
\phi_m(\crh)^{-1} =
{1\over 2} \left\{   \eta\left({\crh_s\over 4}\right)^{-24}
+ \eta\left({\crh_s+2\over 4}\right)^{-24} 
\right\}\, .
\ee
Thus we have
\be \label{ev=0}
\cp(\crh,\cs,\cv) \sim \left[\cv_s^2 \, \left\{   \eta\left({\cs_s
\over 4}
\right)^{-24}
+ \eta\left({\cs_s+2\over 4}\right)^{-24} 
\right\}^{-1}
\left\{   \eta\left({\crh_s\over 4}\right)^{-24}
+ \eta\left({\crh_s+2\over 4}\right)^{-24} 
\right\}^{-1} +\OO(\cv_s^4)\right]\, ,
\ee
near $\cv=0$.
One can easily verify that the functions $\phi_e(\cs)$ and
$\phi_m(\crh)$
transform
as modular forms of weight 12 under the transformation 
$\cs_s\to (p\cs_s + q) / (r\cs_s + s)$ and
$\crh_s\to (\alpha \crh_s +\beta)/(\gamma\crh_s+\delta)$
with $\pmatrix{p & q\cr r & s}
\in \Gamma^0(2)$ and $\pmatrix{\alpha & \beta\cr \gamma &\delta}
\in \Gamma^0(2)$.
These can be regarded as symplectic transformations of the
form
\ben \label{esut1}
&\pmatrix{\alpha & 0 & \beta & 0\cr
0 & 1 & 0 & 0\cr \gamma & 0 & \delta & 0\cr
0 & 0 & 0 & 1} \quad \hbox{and} \quad 
\pmatrix{1 & 0 & 0 & 0\cr
0 & p & 0 & q\cr 0 & 0 & 1 & 0\cr
0 & r & 0 & s}, \nonumber \\
& \alpha\delta - \beta\gamma=1, \quad
ps - qr=1, \quad  p,r,s,\alpha,\gamma,\delta\in\ZZZ, \quad
q,\beta\in 2\ZZZ\, ,
\een
acting on $(\crh_s,\cs_s,\cv_s)$.

Next we consider the wall associated with the decay
$(Q,P)\to ((Q-P)/2, (P-Q)/2) + ((Q+P)/2, (Q+P)/2) $. From
\refb{ede1} we see that the
associated matrix can be taken to be
\be \label{ematrix}
\pmatrix{a_0 & b_0\cr c_0 & d_0} = \pmatrix{{1\over\sqrt 2} & 
{1\over\sqrt 2} \cr - {1\over\sqrt 2} & {1\over\sqrt 2}}\, .
\ee
According to \refb{epole1}
this controls the 
behaviour of $\cp(\crh,\cs,\cv)$ near its zero at
\be \label{emm1}
\crh-\cs = 0\, .
\ee
Following the procedure outlined in 
eqs.\refb{esx3}-\refb{ebe1} we can find the coefficient of
$(\crh-\cs)^2$ in the expression for $\cp$. 
One can see from \refb{esa1} that in this case 
${1\over 2} ((Q+P)/2)^2$
and ${1\over 2} ((Q-P)/2)^2$ can take all possible
independent integer
values $(K+1)$ and $(m-J)$ respectively. 
We find from \refb{ebb1} that the inverses of the 
relevant half-BPS partition functions are:
\be \label{emm2}
\phi_e\left(\tau;a_0,c_0\right)
=\eta(2\tau)^{24}, \quad
\phi_m\left(\tau;b_0,d_0\right)
=\eta(2\tau)^{24}\, .
\ee
The factor of $2$ in the argument of $\eta$ is due to the fact that
$Q^{\prime 2}/2 = (d_0Q - b_0 P)^2/2 =
(Q-P)^2/4$ and $P^{\prime 2}/2
=(-c_0Q + a_0 P)^2/2=(Q+P)^2/4$ entering in
\refb{ebb1} are twice the usual integer
normalized combinations ${1\over 8} (Q\pm P)^2$.
This gives, from \refb{esx3}, \refb{ebe1} and \refb{escale2}
\be \label{emm3}
\cp(\crh,\cs,\cv) \sim \left\{(\crh_s-\cs_s)^2 \,
\eta((\crh_s+\cs_s-2\cv_s)/4)^{24} \, \eta((\crh_s+\cs_s+2\cv_s)/4)^{24}
+\OO((\crh_s-\cs_s)^4)\right\}\, ,
\ee
near $\crh_s\simeq\cs_s$. Since $\eta(2\tau)$ transforms covariantly
under $\tau\to (\alpha\tau+{1\over 2}\beta)/
(2\gamma\tau +\delta)$ with $\pmatrix{\alpha &\beta
\cr \gamma& \delta}\in SL(2,\ZZZ)$,
both $\phi_e$ and $\phi_m$
have full $SL(2,\ZZZ)$ symmetry. Using \refb{egensp},
\refb{esecond} and \refb{esym1}
to represent them as symplectic transformations
on the variables $(\crh_s,\cs_s,\cv_s)$ we get the following two
sets of symplectic matrices:
\ben \label{emm4}
& {1\over 2}
\pmatrix{{\alpha_1+1} & {\alpha_1-1} 
& 2{\beta_1} & 2{\beta_1}\cr
{\alpha_1-1} & {\alpha_1+1} 
& 2{\beta_1} & 2{\beta_1}\cr
{\gamma_1/2} & {\gamma_1/2} &
{\delta_1+1} & {\delta_1-1} \cr
{\gamma_1/2} & {\gamma_1/2} &
{\delta_1-1} & {\delta_1+1}
} , \quad 
{1\over 2}\pmatrix{{p_1+1} & {-p_1+1} 
& {2q_1} & {-2q_1}\cr
{-p_1+1} & {p_1+1} 
& {-2q_1} & {2q_1}\cr
{r_1/2} & {-r_1/2}  & {s_1+1} 
& {-s_1+1}\cr
{-r_1/2} & {r_1/2}  & {-s_1+1} 
& {s_1+1}}\, , \nonumber \\ \cr
& \alpha_1,\beta_1,\gamma_1,\delta_1,p_1, q_1, r_1, s_1\in\ZZZ,
\quad \alpha_1\delta_1-\beta_1\gamma_1=p_1s_1-q_1r_1=1\, .
\een  

Next we turn to the wall corresponding to the decay
$(Q,P)\to (Q-P, 0) +(P, P)$. This corresponds to the
choice 
\be\label{ea0b0}
\pmatrix{a_0 & b_0\cr c_0 & d_0} = \pmatrix{1 & 1 \cr 0 & 1}\, ,
\ee
and is associated with the zero of
$\cp$ at
\be \label{eabab1}
\cs +\cv=0\, .
\ee
Since $(Q-P)^2/8=m-J$ and $P^2/4=K-J$ can take independent
integer values,  we should be able to use \refb{ex3},
\refb{ex4}.
The behaviour of $\cp$ near this zero is however somewhat
ambiguous since one of the decay products -- the state carrying
charge $(Q-P,0)$ -- is not a primitive dyon. As a result the index
associated with this state is ambiguous.\footnote{For half-BPS states
in $\NN=2$ supersymmetric theories a modification of the wall
crossing formula for such non-primitive decays has been suggested
in \cite{0702146}. It is not clear {\it a priori}
how to modify it for the decays of
quarter BPS dyons in $\NN=4$ supersymmetric string theories.
In \S\ref{sprop} we shall propose a formula for the partition function
of the states being studied in this section and examine it to find what the
modification should be.}
Nevertheless if we
go ahead and assume the naive index that follows from
tree level spectrum of elementary string states, we get the
following factorization behaviour of $\cp$:
\be \label{eabab2}
\cp(\crh, \cs, \cv) \, {?\atop \sim} \, \left\{ (\cs_s + \cv_s)^2 
\phi_e\left({\cs_s\over 4};1,0\right) \, 
\phi_m\left({\crh_s+\cs_s+2\cv_s\over 4};1,1\right) 
 +\OO((\cs_s+\cv_s)^4)\right\}\quad
\hbox{for $\cv\simeq -\cs$}\, ,
\ee
with
\ben \label{eabab2.5}
\phi_e\left({\tau\over 4};1,0\right) &=&{1\over 4}
\left\{\eta(\tau/4)^{-24} + \eta((\tau+1)/4)^{-24}
+\eta((\tau+2)/4)^{-24} +\eta((\tau+3)/4)^{-24} 
\right\}^{-1}, \nonumber \\  
\phi_m\left({\tau\over 4};1,1\right) &=&
{1\over 2} \left\{\eta(\tau/4)^{-24} 
+\eta(\tau+2)/4)^{-24}\right\}^{-1}\,  .
\een
$\phi_e(\tau/4)$ has duality symmetries of the form $\tau\to
(p_2\tau+q_2)/(r_2\tau+s_2)$ with 
$\pmatrix{p_2 & q_2\cr r_2 & s_2}\in
\Gamma_0(2)$. On the other hand $\phi_m(\tau/4)$ has duality
symmetries of the form $\tau\to (\alpha_2\tau+\beta_2)/
(\gamma_2\tau+\delta_2)$ with $\pmatrix{\alpha_2 & \beta_2
\cr \gamma_2 &\delta_2}\in \Gamma^0(2)$.
Using \refb{egensp},
\refb{esecond} and \refb{esym1} we find that
the modular properties in this factorized limit correspond to
the following symplectic transformations acting on 
$(\crh_s,\cs_s,\cv_s)$
\ben \label{eabab3}
& \pmatrix{\alpha_2 & \alpha_2 -1 & \beta_2 & 0\cr
0 & 1 & 0 & 0\cr \gamma_2 & \gamma_2 & \delta_2 & 0\cr
\gamma_2  & \gamma_2  & \delta_2 -1 & 1}
\quad \hbox{and} \quad
\pmatrix{1 & 1-p_2 & q_2 & -q_2\cr 0 & p_2 & -q_2 & q_2\cr
0 & 0 & 1 & 0\cr 0 & r_2 & 1-s_2  & s_2}
\, , \nonumber \\
& \alpha_2\delta_2 - \beta_2\gamma_2=1=p_2 s_2-r_2 q_2,
\quad \alpha_2, \gamma_2, \delta_2,
p_2, q_2, s_2 \in\ZZZ, \quad
\beta_2, r_2\in 2\ZZZ \, . 
\een

We can now try to see if all the symplectic transformation
matrices \refb{ecollect}, \refb{ete2s}, \refb{esut1},
\refb{emm4} and \refb{eabab3},
representing possible symmetries of $\cp$, fit into
some subgroup of $Sp(2,\ZZZ)$ defined by some
congruence condition. As it stands there does not seem to be a
simple
congruence subgroup of $Sp(2,\ZZZ)$ 
that fits all the matrices since some of these matrices do not even
have integer entries. However if
we restrict $\gamma$ and $r$ in \refb{esut1} to be even, \i.e.\ assume
that only a $\Gamma(2)\times \Gamma(2)$ subgroup of the symmetry
group $\Gamma^0(2)\times \Gamma^0(2)$ of the $\cv\to 0$ limit
survives as a symmetry of the full partition function, and restrict
$\gamma_1$ and $r_1$ in \refb{emm4} to be multiples of 4, \i.e.\ assume
that only a $\Gamma_0(4)\times \Gamma_0(4)$ subgroup of the
$\crh_s\to \cs_s$ limit survives as a symmetry of the full partition
function,
then there
is a simple congruence subgroup of $Sp(2,\ZZZ)$ into which all the
matrices fit:
\be \label{epropose}
\pmatrix{1+u & u & v & v\cr u & 1+u & v & v\cr
w & w & 1+u & u \cr w & w & u & 1+u}\quad \hbox{mod 2}, \quad
u,v,w=0, 1\, .
\ee
We speculate that this could be the symmetry group of the dyon partition
function under consideration. 

Finally we turn to the constraint from black hole entropy. As in
\S\ref{s3.1}, in this case we have $g(\tau)=\eta(\tau)^{24}$
in \refb{eblack4}.  Thus \refb{eblack6} takes the
form
\be \label{eblack6bb}
\cp(\crh, \cs,\cv) \propto (2v-\rho-\sigma)^{10}\,
\{ v^2 \, \eta(\rho)^{24}\, 
\eta(\sigma)^{24} +  \OO(v^4)\}\, ,
\ee
where $(\rho,\sigma,v)$ and $(\crh,\cs,\cv)$ are related via
\refb{eblack7}. 

Before concluding this section we would like to note that we can
easily extend the analysis of this section to the complementary subset
of torsion 2 dyons with $Q$, $P$ primitive and $Q^2/2$ and $P^2/2$
odd. For this we consider six dimensional electric and magnetic charge
vectors with metric
\be \label{esix1}
L=\pmatrix{ 0 & 1 & 0 & 0\cr 1 & 0 & 0 & 0\cr 0 & 0 & 0 & I_2\cr
0 & 0 & I_2 & 0}\, ,
\ee
and take the set $\AAA$ to be the collection of charge vectors $(Q,P)$
with
\be \label{esix2}
Q=\pmatrix{1\cr -1\cr 1 \cr 2m+1\cr 1 \cr 1}\, , \qquad P = 
\pmatrix{1\cr 1\cr 2K+1 \cr 2J+1 \cr 1\cr -1}\, , \quad m, K, J\in\ZZZ\, .
\ee
This has
\be \label{esix3}
{Q^2\over 2} = 2m+1, \quad {P^2\over 2} = 2(K-J)+1, 
\quad Q\cdot P = 2(K+J-m+1)\, .
\ee
Thus we have $Q^2/2$ and $P^2/2$ odd and $Q\cdot P$ even. 
Furthermore
we still have the constraint that 
$Q^2 + P^2 + 2Q\cdot P$ is a multiple of 8. Thus with 
$(\crh_s,\cs_s,\cv_s)$ defined as in \refb{escale2}, the partition
function is antiperiodic under $(\crh_s,\cs_s,\cv_s)\to
(\crh_s+2, \cv_s,\cs_s), (\crh_s,\cs_s+2,\cv_s)$ and periodic
under $(\crh_s,\cs_s,\cv_s)\to (\crh_s,\cs_s,\cv_s+2),
(\crh_s+1,\cs_s+1,\cv_s+1)$. We can now repeat the analysis of this
section for this set of dyons. The results are more or less identical
except for some 
relative minus signs between the terms in the curly brackets
in
eqs.\refb{ex11}-\refb{ev=0} and the second equation
in \refb{eabab2.5}.

\subsectiono{Dyons in $\ZZZ_2$ CHL orbifold with twisted sector
electric charge} \label{s3.3}

We now consider a $\ZZZ_2$ CHL orbifold 
defined as follows\cite{9505054,9506048}.
We begin with $E_8\times E_8$
heterotic string theory on $T^4\times S^1
\times \wt S^1$ with $S^1$ and $\wt S^1$ labelled by coordinates
with period $4\pi$ and $2\pi$ respectively, and 
take a quotient of the theory
by a $\ZZZ_2$ symmetry that involves $2\pi$ shift along $S^1$
together with an exchange of the two $E_8$ factors.
In the four dimensional subspace of charges
given in \refb{ess1}, now the momentum
$n'$ along $S^1$ is quantized in units of $1/2$ whereas the
Kaluza-Klein monopole charge
$N'$ along $S^1$ is quantized in units of $2$\cite{0605210}. 
We shall take the set $\AAA$ to be
consisting of charge vectors of the form
\be \label{ed1}
Q=\pmatrix{0 \cr m/2\cr 0 \cr -1}\, , \qquad P = 
\pmatrix{K \cr J \cr 1\cr 0}\, , \quad m, K, J\in\ZZZ\, .
\ee
For this state we have
\be \label{ed2}
Q^2 = - m, \quad P^2 = 2 K, \quad Q\cdot P = -J\, .
\ee
As usual we denote by $\BBB$ the set of all $(Q,P)$ which are
related to the ones given in \refb{ed2} by a T-duality transformation.
Since $Q^2/2$, $P^2/2$ and $Q\cdot P$ are quantized in units of
$1/2$, 1 and 1 respectively, $\cp$ satisfies the periodicity
conditions \refb{e14} with
\be \label{ed3}
a_1\in\ZZZ, \quad a_2\in 2\ZZZ, \quad a_3 \in\ZZZ\, .
\ee
Comparison of \refb{ed1} and \refb{ess1} 
shows that the winding charge
$-w'$ along $S^1$ is
$1$ for this state. Thus it represents a twisted sector state.

Our next task is to determine the subgroup of the S-duality group
that leaves the set $\BBB$ invariant. In this case the full S-duality
group is $\Gamma_0(2)$, generated by matrices of the form
$\pmatrix{a & b\cr c & d}$ with $a,b,d\in\ZZZ$, $c\in 2\ZZZ$,
$ad-bc=1$. It was shown in \cite{0708.1270} 
that the set $\BBB$ is closed
under the full S-duality group. Thus the full S-duality group must
be a symmetry of the partition function.

We now turn to the constraints from the wall crossing formula.
Consider first the wall associated with the decay $(Q,P)\to (Q,0)
+(0,P)$, --
this in fact is the only case we need to analyze since 
all the
walls are related to this one by S-duality 
transformation\cite{0702141}. First of all note from
\refb{ed1}, \refb{ed2} that for a given $Q^2=-m$ the charge vector 
$Q\in\AAA$
is fixed uniquely. Thus the index of half-BPS states with charge
$(Q,0)$ can be regarded as a function of $Q^2$. On the other hand
for a given $P^2=2K$ there is a family of $P\in\AAA$ labelled by
$J$, but  these can be transformed to the vector corresponding to
$J=0$ by the T-duality transformation matrix\cite{0708.1270}
$\pmatrix{1 & 0 & 0 & J\cr 0 & 1 & -J & 0\cr
0 & 0 & 1 & 0\cr 0 & 0 & 0 & 1}$. Thus the index of the charge
vector $(0,P)$ can also be expressed as a function of $P^2$.
Finally we see from \refb{ed2} that the allowed values of $Q^2$
and $P^2$ are uncorrelated. Thus we can use eqs.\refb{ex3},
\refb{ex4} to extract the behaviour of $\cp$ near $\cv=0$.
The electric partition function can be calculated by
examining the spectrum of twisted sector states in the heterotic
string theory\cite{0502126,0502157,0504005,0507014}. 
On the other hand the magnetic partition function
can be calculated by examing the spectrum of D1-D5 system in a dual
type IIB description of the theory\cite{0708.1270}. The results are
\be \label{ed4}
\phi_e(\cs) = \eta(\cs)^{8} \eta(\cs/2)^{8}\, ,
\qquad \phi_m(\crh)  = \eta(\crh)^{8} \eta(2\crh)^{8}\, .
\ee
Eq.\refb{ex3} then gives, near $\cv=0$,
\be \label{efactor3}
\cp(\crh,\cs,\cv) \propto \left\{ \cv^2 \eta(\cs)^8
\eta(\cs/2)^8 \eta(\crh)^8\eta(2\crh)^8 +\OO(\cv^4)\right\}\, .
\ee
$\phi_e(\cs)$ and $\phi_m(\crh)$ 
transform as modular forms of weight $8$ under
\be \label{ed5}
\cs\to {p\cs + q\over r\cs +s}, \quad p,r,s\in\ZZZ, \quad q\in 2\ZZZ,
\quad ps - qr=1\, ,
\ee
and
\be \label{ed6}
\crh\to {\alpha \crh +\beta\over \gamma\crh+\delta}, \quad
\alpha,\beta,\delta\in \ZZZ, \quad \gamma\in 2\ZZZ, \quad
\alpha\delta - \beta\gamma=1\, .
\ee
The corresponding groups are $\Gamma^0(2)$ and
$\Gamma_0(2)$ respectively. Thus from \refb{e13},
\refb{e15}, \refb{ez1}
we see that if \refb{ed5} and \refb{ed6} lift to symmetries
of the full partition function then the partition
function transforms as a modular form of weight 6 under the
$Sp(2,\ZZZ)$ transformations of the form 
\be \label{es1b}
\pmatrix{d & b &0&0 \cr c & a &0&0\cr
0&0& a & -c\cr 0&0& -b & d}, \quad 
\pmatrix{1 & 0 & a_1 & a_3\cr
0 & 1 & a_3 & a_2\cr 0 & 0 & 1 & 0\cr 0 & 0 & 0 & 1}, \quad
\pmatrix{\alpha & 0 & \beta & 0\cr
0 & 1 & 0 & 0\cr \gamma & 0 & \delta & 0\cr
0 & 0 & 0 & 1} \quad \hbox{and} \quad 
\pmatrix{1 & 0 & 0 & 0\cr
0 & p & 0 & q\cr 0 & 0 & 1 & 0\cr
0 & r & 0 & s}\, ,
\ee
with
\be \label{ed7}
\pmatrix{a & b\cr c & d}\in \Gamma_0(2), \quad
\pmatrix{\alpha & \beta\cr \gamma & \delta}
\in \Gamma_0(2), \quad \pmatrix{p & q\cr r & s} \in
\Gamma^0(2), \quad a_1, a_3\in \ZZZ, \quad a_2 \in 2\ZZZ\,.
\ee
All the $Sp(2,\ZZZ)$ matrices in \refb{es1b}
subject to the constraints \refb{ed7} have the form
\be \label{edd8}
\pmatrix{1 & * & * & *\cr 0 & 1 & * & 0\cr 0 & 0 & 1 & 0\cr
0 & * & * & 1} \quad \hbox{mod 2}\, .
\ee
Furthermore the set of matrices \refb{edd8} are closed under
matrix multiplication. Thus 
the group generated by the set of $Sp(2,\ZZZ)$
matrices \refb{es1b} subject to the
condition \refb{ed7} is contained in the group
$\check G$ of $Sp(2,\ZZZ)$ matrices \refb{edd8}.

All the symmetries listed in \refb{es1b} 
are indeed symmetries of the dyon partition function of
this model proposed in \cite{0510147} and 
proved in \cite{0605210}. 
Furthermore near $\cv=0$ the partition function is known to
have the factorization property given in 
\refb{efactor3}\cite{0702141,0705.3874,0706.2363}.
One question that one can ask is: do
the matrices given in \refb{es1b} generate the full symmetry
group of the partition function (which is known in this case)? It
turns out that the answer is no. This group does not include the
matrix
\be \label{emiss}
\pmatrix{1 & 0 & 0 & 0\cr 0 & 1 & 0 & 0\cr 0 & -1 & 1 & 0\cr
-1 & 0 & 0 & 1}\, ,
\ee
since this is not of the form given in \refb{edd8}. This
generates the transformation
\be \label{emissym}
\crh \to {\crh\over (1 - \cv)^2 - \cs\crh}, \quad
\cs \to {\cs \over (1 - \cv)^2 - \cs\crh}, \quad
\cv \to {\crh \cs + \cv(1 - \cv)
 \over (1 - \cv)^2 - \cs\crh}\, ,
 \ee
 and is known to be a symmetry of the partition 
 function.\footnote{This
 is the symmetry refered to as $g_3(1,0)$ in \cite{0510147} in a 
 different representation.}

Finally we turn to the constraints from black hole entropy.
In this case the function $g(\tau)$ is given 
by\cite{9708062,0502126}:
\be \label{egtau}
g(\tau) = 
\eta(\tau)^{8}\, \eta(2\tau)^{8}\, .
\ee
Thus \refb{eblack6} takes the
fom
\be \label{eblack6c}
\cp(\crh, \cs,\cv) \propto (2v-\rho-\sigma)^{6}\,
 \{ v^2 \, \eta(\rho)^{8}\eta(2\rho)^8\, 
\eta(\sigma)^{8}\eta(2\sigma)^8 +  \OO(v^4)\}\, ,
\ee
where $(\rho,\sigma,v)$ and $(\crh,\cs,\cv)$ are 
related via \refb{eblack7}.
The dyon partition function of the $\ZZZ_2$ CHL model is
known to satisfy this property. In fact
historically this is the property that
was used to guess the form of the partition function\cite{0510147}.

This analysis can be easily generalized to the dyons of $\ZZZ_N$
CHL orbifolds carrying twisted sector electric charges.

\subsectiono{Dyons in $\ZZZ_2$ CHL model 
with untwisted sector
electric charge} \label{s3.5}

We again consider the $\ZZZ_2$ CHL model introduced
in \S\ref{s3.3}, but now take the set $\AAA$ to consist of
dyons with charge vectors
\be \label{ed1a}
Q=\pmatrix{0 \cr (2m+1)/2\cr 0 \cr -2}\, , \qquad P = 
\pmatrix{2K+1 \cr J \cr 1\cr 0}\, , \quad m, K, J\in\ZZZ\, .
\ee
Since $w'=-2$ for this
state, it represents an untwisted sector state.
For this state we have
\be \label{ed2a}
Q^2 = - 2(2m+1), \quad P^2 = 2(2 K+1), \quad Q\cdot P = -2J\, .
\ee
Note that $Q$ and $P$ are both primitive. Since $Q\cdot P$ is
quantized in units of 2, we shall define 
\be \label{esev1}
Q_s={Q\over \sqrt
2}, \quad P_s={P\over \sqrt 2}, \quad  
\crh_s = 2\crh, \quad \cs_s = 2 \cs, \quad \cv_s = 2\cv\, .
\ee
Thus we have
\be \label{ef1}
Q_s^2 = -\left(2m+1\right), \quad P_s^2 = 2K+1, \quad
Q_s\cdot P_s=-J\, .
\ee
Since $Q_s^2/2$ is quantized in units of 1/2, we expect the partition
function to have $\cs_s$ period $2$. However except for an
overall additive factor of $1/2$, $Q_s^2/2$ is actually quantized in
integer units. Thus the partition function has the additional
property that it is odd under $\cs_s\to \cs_s + 1$. 
Similarly since $P_s^2$ is an odd integer, the partition function
picks up a minus sign under $\crh_s\to\crh_s+1$. We shall call these
symmetries of $\cp$. Finally
since  $Q_s\cdot P_s$
is quantized in 
integer units,  the period in the  $\cv_s$
direction is also unity. 
The corresponding symplectic transformations acting on
$(\crh_s,\cs_s,\cv_s)$ are of the form
\be \label{espsp1}
\pmatrix{1 & 0 & \wt a_1 & \wt a_3\cr 0 & 1 & \wt a_3 & 
\wt a_2\cr
0 & 0 & 1 & 0\cr 0 & 0 & 0 & 1}\, , \quad \wt a_1,
\wt a_2, \wt a_3\in\ZZZ\, .
\ee
Under this transformation the partition function picks up a 
multiplier factor 
of $(-1)^{\wt a_1+\wt a_2}$.

Our next task is to determine the subgroup of the S-duality group
$\Gamma_0(2)$ that leaves the set $\BBB$ -- defined as the
T-duality orbit of $\AAA$ -- invariant. 
For this let us apply the S-duality transformation 
$\pmatrix{a & b\cr c & d}\in\Gamma_0(2)$ 
on the charge vector \refb{ed1a}.
This gives
\be \label{ef2}
Q' = aQ + bP = \pmatrix{b (2K+1)\cr (2m+1)a/2 + bJ \cr b \cr -2a}, 
\qquad P'=cQ + dP =\pmatrix{d (2K+1)
\cr (2m+1)c/2 + dJ \cr d \cr -2c}
\, .
\ee
We need to choose $a$, $b$, $c$, $d$ such that \refb{ef2} is
inside the set $\BBB$, \i.e. it can be brought to the form  
\refb{ed1a} after a T-duality transformation. The T-duality
transformations acting within this four dimensional subspace
are generated by matrices of the form\cite{0708.1270}:
\be \label{ef3}
\pmatrix{n_1 & -m_1 && \cr -l_1 & k_1 && \cr && k_1 & l_1\cr
&& m_1 & n_1} \quad \hbox{and} \quad
\pmatrix{k_2 &&& -l_2\cr & k_2&l_2&\cr &m_2&n_2&\cr
-m_2&&&n_2}, \quad   \pmatrix{k_i & l_i\cr m_i & n_i}
\in \Gamma_0(2)\, .
\ee
Now suppose $b$ in \refb{ef2} is even. Then we can apply a
T-duality transformation on the charge vector given in
\refb{ef2} with the matrix
\be \label{ef4}
\pmatrix{1 && &l_0\cr &1& -l_0 &\cr 
& 0 & 1 &\cr 0 &&& 1} 
\pmatrix{d & -2c && \cr -b/2 & a &&\cr && a & b/2\cr
&& 2c & d}, \qquad l_0 \equiv {1\over 2} bd(2K+1)
-{c\over 2} \{ (2m+1) a + 2 b J\}\, .
\ee
It is straightfoward to verify that this brings \refb{ef2}
back to the set $\AAA$ consisting of pairs of charge
vectors of the form given in \refb{ed1a}. 
This shows that a sufficient condition for \refb{ef2} to
lie in the set $\BBB$ is to have $b$ even, \i.e. 
$\pmatrix{a & b\cr c & d}\in\Gamma(2)$. Using
\refb{e3.5} we can also see that this condition is necessary
since acting on a pair $(Q,P)$ with $Q^2/2$ odd, $P^2/2$
odd and $Q\cdot P$ even, an S-duality transformation produces
a $(Q',P')$ with odd $Q^{\prime 2}/2$ only if $b$ is even.

Thus we identify
the subgroup $\Gamma(2)$ of the S-duality group 
$\Gamma_0(2)$ as the symmetry of the set $\BBB$. The overall
scaling of $Q$ and $P$ does not change the symmetry group.
Thus the
quarter BPS dyon partition function associated with the set $\BBB$
must be invariant under the $\Gamma(2)$ S-duality symmetry.
This in turn corresponds to symplectic transformations of the
form
\be \label{espsp3}
\pmatrix{d & b &0&0 \cr c & a &0&0\cr
0&0& a & -c\cr 0&0& -b & d}, \qquad ad-bc=1,\quad 
a,d\in\ZZZ, 
\quad b,c \in 2\ZZZ\, ,
\ee
acting on $(\crh,\cs,\cv)$ and also on $(\crh_s,\cs_s,\cv_s)$.

Next we turn to the analysis of the constraints from
wall crossing. First consider the wall
corresponding to the decay $(Q,P)\to (Q,0)+(0,P)$, and examine
whether there are subtleties of kind mentioned below eq.\refb{ex5}
in applying eqs.\refb{ex3}, \refb{ex4}. For this we note that
here $Q^2=-2(2m+1)$ and $P^2=2(2K+1)$ are uncorrelated.
For a given $Q^2= -2(2m+1)$ there is a unique charge vector
in the list given in \refb{ed1a}. On the other hand even though for
a given $P^2=2(2K+1)$ there is an infinite family of $P$ labelled
by $J$, they are all related by the T-duality transformation
matrix $\pmatrix{1&0 & 0 &-J\cr 0 & 1 & J & 0\cr 
0 & 0 & 1 & 0\cr
0 & 0 & 0 & 1}$ to the vector $\pmatrix{2K+1\cr 0\cr 1\cr 0}$.
Thus there are no subtleties of the kind mentioned below
\refb{ex5} and we have
\be \label{espsp4}
\cp(\crh,\cs,\cv) \propto \left\{ \cv^2 \, \phi_m(\crh)
\phi_e(\cs) +\OO(\cv^4)\right\} 
\quad \hbox{for $\cv\simeq 0$}\, .
\ee
The magnetic partition function
is obtained from \refb{ed4} by  projection
to odd values of $P^2/2$ followed by
$\crh\to\crh_s/2$
replacement.  This
gives
\be \label{epm1}
\phi_m(\crh)^{-1} = {1\over 2}
\left\{ \eta(\crh_s/2)^{-8} \eta(\crh_s)^{-8}
- \eta((\crh_s+1)/2)^{-8} \eta(\crh_s)^{-8}\right\}\, .
\ee
On the other hand the electric partition function can be
calculated by analyzing the untwisted sector BPS spectrum
of the fundamental heterotic 
string\cite{0502126,0502157,0504005,0507014}. After taking into
account the fact that we are
computing the partition function of odd $Q^2/2$ states
only, and the $\cs\to \cs_s/2$ replacement, the result is
\ben \label{epm2}
\phi_e^{-1}(\cs) &=& {1\over 2} \left( \psi_e(\cs_s) - \psi_e
\left(\cs_s+1\right)\right) \, , \nonumber \\
\psi_e(\cs_s) &=&
8\, \eta(\cs_s/2)^{-24}\, \left[{1\over 2} \left(
\vt_2(\cs_s)^8 + \vt_3(\cs_s)^8 + \vt_4(\cs_s)^8\right)
- \vt_3(\cs_s/2)^4 \vt_4(\cs_s/2)^4
\right] \, . \nonumber \\
\een
In \refb{epm2} $\psi_e$ describes the partition function
before projecting on to the odd $Q^2/2$ sector\cite{0502157}. 

$\phi_m(\crh_s)$ given in \refb{epm1}
transforms as a modular form of weight 8 under 
$\crh_s\to (\alpha\crh_s+\beta)/(\gamma\crh_s+\delta)$ with
$\pmatrix{\alpha &\beta\cr \gamma &\delta}\in \Gamma_0(2)$
with a multiplier $(-1)^\beta$. On the other hand
$\phi_e(\cs)$ given in \refb{epm2} can be shown to transform as
a modular form of weight $8$ under 
$\cs_s\to (p\cs_s + q)/(r \cs_s +s)$
for $\pmatrix{p & q\cr r & s} \in \Gamma_0(2)$, with a
multiplier $(-1)^q$.
These duality symmetries correspond to the symplectic 
transformations
\ben \label{es1d}
&\pmatrix{\alpha & 0 & \beta & 0\cr
0 & 1 & 0 & 0\cr \gamma & 0 & \delta & 0\cr
0 & 0 & 0 & 1} \quad \hbox{and} \quad 
\pmatrix{1 & 0 & 0 & 0\cr
0 & p & 0 & q\cr 0 & 0 & 1 & 0\cr
0 & r & 0 & s}\, , \nonumber \\ \cr
& \alpha\delta - \beta\gamma=1, \quad ps-qr=1, \quad
\alpha, \beta, \delta, p, q, s\in \ZZZ, 
\quad \gamma, r\in 2\ZZZ\, ,
\een
acting on $(\crh_s,\cs_s,\cv_s)$.

Since for the set $\BBB$ the S-duality group is $\Gamma(2)$,
in this case there is another wall of marginal stability, associated
with $\pmatrix{a_0 & b_0\cr c_0 & d_0}
=\pmatrix{1 & 1\cr 0 & 1}$,
which cannot be related to the previous wall by an S-duality
transformation. This corresponds to the decay $(Q,P)\to (Q-P, 0)
+ (P,P)$ and controls the behaviour of $\cp$ near $\cv+\cs=0$. 
As usual we first need to determine if there are any subtleties
of the type mentioned below eq.\refb{ex5}. Eq.\refb{ed1a} shows that
for a given $(Q-P)^2=4(K+J-m)$ there is an infinite family of
$(Q-P)=\pmatrix{-(2K+1)\cr (2m-2J+1)/2\cr -1\cr -2}$ labelled
by $2K$. However all of these can be related by T-duality
transformation $\pmatrix{1 & 0 & 0 & K\cr 0 & 1 & -K & 0\cr
0 & 0 & 1 &0\cr 0 & 0 & 0 & 1}$ to the vector
$\pmatrix{-1\cr (2m-2J-2K+1)/2\cr -1\cr -2}$ which is
determined completely in terms of $(Q-P)^2$. We have already
seen earlier that all choices of $P$ for a given $P^2$ are
also related by  T-duality transformations. Finally we note that
in this case $(Q-P)^2/2$ and $P^2/2$ can take independent even
and odd integer values respectively. It then follows that there
are no subtleties of the kind mentioned 
below \refb{ex5}. After evaluating $\phi_e$ and $\phi_m$
by standard procedure we find that near $\cv+\cs=0$ $\cp$
behaves as
\be \label{efin1}
\cp(\crh,\cs, \cv) \propto \left[
\cv_s^{\prime 2} \, \left\{ \eta(\crh_s'/2)^{-8}
- \eta((\crh_s'+1)/2)^{-8}
\right\}^{-1} \eta(\crh_s')^{8} \{ \psi_e(\cs_s') + \psi_e(\cs_s'+1) \}^{-1}
+\OO(\cv_s^{\prime 4})
\right]\, ,
\ee  
where
\be \label{efin2}
\cv_s'= \cv_s+\cs_s, \quad \cs_s' = \cs_s, 
\quad \crh_s' = \crh_s +\cs_s+2\cv_s\, .
\ee
This has duality symmetry
$\crh_s'\to (\alpha_1\crh_s'+\beta_1)/(\gamma_1\crh_s' +\delta_1)$  and
$\cs_s'\to (p_1\cs_s' + q_1)/(r_1\cs_s' + s_1)$ for
$\pmatrix{\alpha_1 & \beta_1\cr \gamma_1 & \delta_1}
\in \Gamma_0(2)$
and $\pmatrix{p_1 & q_1\cr r_1 & s_1}\in \Gamma_0(2)$
with a multiplier $(-1)^{\beta_1}$. 
We can express them as symplectic transformations acting on
$(\crh_s,\cs_s,\cv_s)$ using
\refb{egensp}, \refb{esecond}. This gives
\ben \label{efin3}
& \pmatrix{\alpha_1 & \alpha_1 -1 & \beta_1 & 0\cr
0 & 1 & 0 & 0\cr \gamma_1 & \gamma_1 & \delta_1 & 0\cr
\gamma_1 & \gamma_1 
& \delta_1 -1 & 1}, \quad \hbox{and} \quad 
\pmatrix{1 & 1-p_1 & q_1 & -q_1\cr 0 & p_1 & -q_1 & q_1\cr
0 & 0 & 1 & 0\cr 0 & r_1 & 1-s_1  & s_1}
\, , \nonumber \\ \cr
& \alpha_1\delta_1 - \beta_1\gamma_1=1=p_1 s_1-r_1 q_1,
\quad \alpha_1, \beta_1, \delta_1,
p_1, q_1, s_1 \in\ZZZ, \quad
\gamma_1, r_1\in 2\ZZZ \, . 
\een

As usual, we would like to know if there is a natural subgroup
of $Sp(2,\ZZZ)$ defined by some congruence relation into which
all the $Sp(2,\ZZZ)$ matrices \refb{espsp1}, \refb{espsp3},
\refb{es1d} and \refb{efin3} fit. There is indeed such a subgroup
defined as the collection of matrices of the form
\be \label{epm3.2}
\pmatrix{1 & 0 & * & *\cr 0 & 1 & * & *\cr
0 & 0 & 1 & 0\cr 0 & 0 & 0 & 1} \quad \hbox{mod 2}\, .
\ee
It is natural to speculate that \refb{epm3.2} is the symmetry group
of the partition function under consideration.

Finally we turn to the constraint from black hole entropy. Since
we are considering $\ZZZ_2$ CHL orbifold, the
function $g(\tau)$ appearing in the coefficient of the Gauss-Bonnet
term in the effective action is the same as the one in
\S\ref{s3.3}:
\be \label{egtaunew}
g(\tau) = 
\eta(\tau)^{8}\, \eta(2\tau)^{8}\, .
\ee
Thus \refb{eblack6} takes the
fom
\be \label{eblack6e}
\cp(\crh, \cs,\cv) \propto (2v-\rho-\sigma)^6\,
\{ v^2 \, \eta(\rho)^{8}\eta(2\rho)^8\, 
\eta(\sigma)^{8}\eta(2\sigma)^8 +  \OO(v^4)\}\, ,
\ee
where $(\rho,\sigma,v)$ and $(\crh,\cs,\cv)$ are related via
\refb{eblack7}.

\renewcommand{\theequation}{\thesection.\arabic{equation}}

\sectiono{A proposal for the
partition function of dyons of torsion two} \label{sprop}

In this section we shall consider the set of dyons described in
\S\ref{s3.4}, carrying charge vectors $(Q,P)$ 
with torsion 2, $Q$, $P$ primitive and $Q^2/2$, $P^2/2$ even,
and propose a form of the partition function that satisfies all the
constraints derived in \S\ref{s3.4}. The proposed form of
the partition function is
\ben \label{epr1}
{1\over \cp(\crh,\cs,\cv)} &=& {1\over 16} \Bigg[
{1\over \Phi_{10}(\crh,\cs,\cv)} +
{1\over \Phi_{10}(\crh,\cs+{1\over 2},\cv)} +
{1\over \Phi_{10}(\crh+{1\over 2},\cs,\cv)} \nonumber \\
&& +
{1\over \Phi_{10}(\crh+{1\over 2},\cs+{1\over 2},\cv)} +
{1\over \Phi_{10}(\crh+{1\over 4},\cs+{1\over 4},\cv+{1\over 4})} \nonumber \\
&&
+
{1\over \Phi_{10}(\crh+{1\over 4},\cs+{3\over 4},\cv+{1\over 4})} +
{1\over \Phi_{10}(\crh+{3\over 4},\cs+{1\over 4},\cv+{1\over 4})} \nonumber \\
&&+
{1\over \Phi_{10}(\crh+{3\over 4},\cs+{3\over 4},\cv+{1\over 4})} 
 +
{1\over \Phi_{10}(\crh+{1\over 2},\cs+{1\over 2},\cv+{1\over 2})} \nonumber \\
&&+
{1\over \Phi_{10}(\crh+{1\over 2},\cs,\cv+{1\over 2})} +
{1\over \Phi_{10}(\crh,\cs+{1\over 2},\cv+{1\over 2})} +
{1\over \Phi_{10}(\crh,\cs,\cv+{1\over 2})}  \nonumber \\
&&+
{1\over \Phi_{10}(\crh+{3\over 4},\cs+{3\over 4},\cv+{3\over 4})} +
{1\over \Phi_{10}(\crh+{3\over 4},\cs+{1\over 4},\cv+{3\over 4})} 
\nonumber \\
&& +
{1\over \Phi_{10}(\crh+{1\over 4},\cs+{3\over 4},\cv+{3\over 4})}  +
{1\over \Phi_{10}(\crh+{1\over 4},\cs+{1\over 4},\cv+{3\over 4})}\Bigg]
\nonumber \\
&&  +  
\Bigg[{1\over \Phi_{10}(\crh+\cs+2\cv,
\crh+\cs-2\cv,\cs-\crh)} \nonumber \\
&&
+ {1\over \Phi_{10}(\crh+\cs+2\cv+{1\over 2},
\crh+\cs-2\cv+{1\over 2},\cs-\crh+{1\over 2})}
\Bigg]\, .
\een
The index $d(Q,P)$ is computed from this partition function using
the formula
\ben \label{epr2}
d(Q,P)&=& {1\over 4}\, (-1)^{Q\cdot P+1} \, \int_\CC
 d\crh_s
d\cs_s  d\cv_s \,
e^{-i{\pi\over 4} (\cs_s Q^2 + \crh_s P^2 
+ 2 \cv_s Q\cdot P)} \, 
{1\over \cp(\crh, \cs, \cv)}\, , \nonumber \\ && \qquad \qquad
(\crh_s,\cs_s,\cv_s) \equiv (4\crh,4\cs,4\cv)\, ,
\een
where the contour $\CC$ is defined by fixing the imaginary
parts of $\crh_s$, $\cs_s$, $\cv_s$ to appropriate values depending on
the domain of the moduli space in which we want to compute the
index, and the real parts span the unit cell defined by the
periodicity condition
\be \label{epr3}
(\crh_s, \cs_s, \cv_s) \to (\crh_s+2, \cs_s, \cv_s), \,
(\crh_s, \cs_s+2, \cv_s), \, (\crh_s, \cs_s, \cv_s+2), 
\, (\crh_s+1, \cs_s+1, \cv_s+1)\, .
\ee
The overall 
multiplicative factor of 1/4 in \refb{epr2} accounts for the
fact that the unit cell defined by eqs.\refb{epr3} has volume 4.
The factor of $1/4$ in the exponent in \refb{epr2}
accounts for the replacement of $(\crh,\cs,\cv)$ by
$(\crh_s/4,\cs_s/4,\cv_s/4)$ in \refb{e2}.
Note that the sixteen terms inside the first square bracket in
\refb{epr1} together gives the partition function of dyons of unit
torsion subject to the constraint that $Q^2/2$, $P^2/2$, $Q\cdot P$
are even and $Q^2+P^2-2Q\cdot P$ is a multiple of 8. The second
term is new and reflects the effect of considering states with torsion
two.

We shall now check that this formula satisfies all the constraints
derived in \S\ref{s3.4}. We begin with the 
S-duality transformations.
The first term inside the first
square bracket $(\Phi_{10}(\crh,\cs,\cv))^{-1}$ is 
S-duality invariant since S-duality transformation can be
regarded as an $Sp(2,\ZZZ)$ transformation on 
$(\crh,\cs,\cv)$. The
other terms inside the first square bracket have the form
$(\Phi_{10}(\crh+b_1,\cs+b_2,\cv+b_3))^{-1}$
for appropriate choices of $(b_1,b_2,b_3)$. Since these shifts
may be
represented as symplectic transformations of $(\crh,\cs,\cv)$,
S-duality transformation \refb{ete2s} will change this term to
$(\Phi_{10}(\crh+b_1'',\cs+b_2'',\cv+b_3''))^{-1}$
with
\ben \label{epr4}
&& \pmatrix{1 & 0 & b_1'' & b_3''\cr
0 & 1 & b_3'' & b_2''\cr 0 & 0 & 1 & 0\cr 0 & 0 & 0 & 1}
=\pmatrix{d & b &0&0 \cr c & a &0&0\cr
0&0& a & -c\cr 0&0& -b & d}^{-1} \, 
\pmatrix{1 & 0 & b_1 & b_3\cr
0 & 1 & b_3 & b_2\cr 0 & 0 & 1 & 0\cr 0 & 0 & 0 & 1}
\pmatrix{d & b &0&0 \cr c & a &0&0\cr
0&0& a & -c\cr 0&0& -b & d}\, ,
\nonumber \\ &&
\qquad a,b,c,d\in\ZZZ, \quad ad-bc=1,
\quad a+c\in 2\ZZZ+1, \quad b+d\in 2\ZZZ+1\, .
\een
One finds that under such a transformation the sixteen triplets
$(b_1,b_2,b_3)$ appearing in the sixteen terms inside the first
square bracket get permuted up to integer shifts which are
symmetries of $\Phi_{10}$. This proves the S-duality invariance
of the first 16 terms.

To test the S-duality invariance of the second term we
define
\be \label{epr5}
\crh' = (\crh+\cs+2\cv), \quad \cs'=
(\crh+\cs-2\cv), \quad \cv' =(\cs-\crh)\, .
\ee
It is easy to verify that the effect of S-duality transformation 
\refb{ete2s} on the
$(\crh',\cs',\cv')$ variables is represented by the symplectic
matrix
\be \label{epr6}
\pmatrix{(a+b+c+d)/2 & (a+b-c-d)/2 &0&0 \cr(a-b+c-d)/2
 & (a-b-c+d)/2 &0&0\cr
0&0& (a-b-c+d)/2 & -(a-b+c-d)/2\cr 0&0& -(a+b-c-d)/2 & 
(a+b+c+d)/2}\,.
\ee
Given the conditions \refb{epr4} on $a,b,c,d$, this is an
$Sp(2,\ZZZ)$
transformation. Thus the first term inside the second 
square bracket in \refb{epr1},
given by $(\Phi_{10}(\crh',\cs',\cv'))^{-1}$, is
manifestly S-duality invariant. The second term involves a
shift of $(\crh',\cs',\cv')$ by $(1/2,1/2,1/2)$. One can easily
check that this commutes with the symplectic transformation
\refb{epr6} up to integer shifts in $(\crh',\cs',\cv')$. Thus the second
term is also S-duality invariant.

We now turn to the wall crossing formul\ae. First consider the wall
associated with the decay $(Q,P)\to (Q,0)+(0,P)$. The jump in
the index across this wall is controlled
by the residue of the pole at $\cv=0$. In order to evaluate this residue
it will be most convenient to choose the unit cell over which
the integration \refb{epr2} is performed to be
$-1\le\Re(\crh_s)<1$, $-1\le\Re(\cs_s)<1$ and $-{1\over 2}\le \Re(\cv_s)
<{1\over 2}$ so that the image of the
pole at $\cv_s=0$ under $(\crh_s,\cs_s,\cv_s)\to 
(\crh_s\pm 1,\cs_s\pm 1,\cv_s\pm 1)$
is outside the unit cell, -- otherwise we would need to include the
contribution from this pole as well.
Using \refb{epr2}
we see that the change in the index across this wall is given by
\be \label{ech1}
\Delta d(Q,P) =
{1\over 4}\, (-1)^{Q\cdot P+1} \, \int_{iM_1-1}^{iM_1+1} 
d\crh_s \int_{iM_2-1}^{iM_2+1} d\cs_s 
\ointop d\cv_s 
 \,
e^{-i{\pi\over 4} (\cs_s Q^2 + \crh_s P^2 
+ 2 \cv_s Q\cdot P)} \, 
{1\over \cp(\crh, \cs, \cv)}\, ,
\ee
where $\ointop$ denotes the contour around $\cv_s=0$ and $M_1$,
$M_2$ are large positive numbers.
Now the poles in \refb{epr1} can be found from the known
locations of the zeroes in $\Phi_{10}(x,y,z)$:
\ben \label{eapl5a}
&& n_2 ( xy-z^2) + j z  + 
n_1 y  -  m_1 x+ m_2 = 0 \nonumber \\
&& m_1,
n_1, m_2, n_2 \in \ZZZ, \quad j\in 2\ZZZ+1, \quad
m_1 n_1 + m_2 n_2 +\frac{j^2}{4} = {1\over 4}\, .
\een
Using this we find that the poles in \refb{epr1} at $\cv_s=0$
can come from the first
four terms inside the first square bracket. There
is no pole at $\cv_s=0$ from the terms in the second square bracket.
The residue at the pole can be calculated by using the fact that
\be \label{ech2}
\Phi_{10}(x,y,z) \simeq -4\pi^2\, z^2\, \eta(x)^{24} 
\eta(y)^{24} + \OO(z^4)\, ,
\ee
near $z=0$. This gives
\be \label{ech2.5}
\cp(\crh,\cs,\cv)
= -4\pi^2\,  \cv_s^2 \, \left\{   \eta\left({\cs_s\over 4}
\right)^{-24}
+ \eta\left({\cs_s+2\over 4}\right)^{-24} 
\right\}^{-1}
\left\{   \eta\left({\crh_s\over 4}\right)^{-24}
+ \eta\left({\crh_s+2\over 4}\right)^{-24} 
\right\}^{-1} +\OO(\cv_s^4) \, .
\ee
Substituting this into \refb{ech1} and using the convention that the
$\cv_s$ contour encloses the pole clockwise, we get
\ben \label{ech3}
\Delta d(Q,P) &=&
{1\over 16}\, (-1)^{Q\cdot P+1} \, Q\cdot P\, 
\int_{iM_1-1}^{iM_1+1} 
d\crh_s \,
\left\{\eta(\crh_s/4)^{-24} + \eta((\crh_s+2)/4)^{-24}\right\}
\, e^{-i\pi \crh_s \, P^2/4}\, \nonumber \\ && 
\int_{iM_2-1}^{iM_2+1} d\cs_s  
\, \left\{\eta(\cs_s/4)^{-24} + \eta((\cs_s+2)/4)^{-24}\right\}
\, e^{-i\pi \cs_s \, Q^2/4} 
\, .
\een
We now want to compare this with the general wall crossing
formula given in \refb{ex1}.
Here the relevant half-BPS partition functions are to be
computed with $Q^2$ and $P^2$ restricted to be even.
This gives
\ben \label{ech4}
d_h(Q,0) &=& \int_{iM-1/4}^{iM+1/4}\, d\tau \,
\left\{ \eta(\tau)^{-24} + \eta\left(\tau+{1\over 2}\right)^{-24}
\right\} \,
e^{-i\pi\tau Q^2}\nonumber \\
& = &
{1\over 4} \,
\int_{4iM-1}^{4iM+1}\,  
d\cs_s \, \left\{\eta(\cs_s/4)^{-24} + \eta((\cs_s+2)/4)^{-24}\right\}
\, e^{-i\pi \cs_s \, Q^2/4}, \nonumber \\
d_h(0,P) &=& \int_{iM-1/4}^{iM+1/4} \, d\tau\,
\left\{ \eta(\tau)^{-24} + \eta\left(\tau+{1\over 2}\right)^{-24}
\right\} \,
e^{-i\pi\tau P^2}\nonumber \\
&=& {1\over 4}\, \int_{4iM-1}^{4iM+1}\, d\crh_s\, 
\left\{\eta(\crh_s/4)^{-24} + \eta((\crh_s+2)/4)^{-24}\right\}
\, e^{-i\pi \crh_s \, P^2/4}\, ,
\een
for some large positive number $M$.
Using this we can rewrite \refb{ech3} as
\be \label{ech5}
\Delta d(Q,P) =
(-1)^{Q\cdot P+1} \, Q\cdot P\, d_h(Q,0)\, d_h(0,P)\, ,
\ee
in agreement with the wall crossing formula.

Next we consider the wall associated with the decay
$(Q,P)\to ((Q-P)/2, (P-Q)/2) + ((Q+P)/2, (Q+P)/2)$.
The associated pole of the partition function is at
$\crh=\cs$. {}With the help of \refb{ech2} 
we find that this pole
arises from the first term inside the second square
bracket  in \refb{epr1}, and near this pole
\be \label{edh1}
\cp(\crh,\cs,\cv) = -{\pi^2\over 4} 
(\cs_s-\crh_s)^2\,
\eta((\crh_s+\cs_s+2\cv_s)/4)\,
\eta((\crh_s+\cs_s-2\cv_s)/4)+\OO\left( (\cs_s-\crh_s)^4\right)\, .
\ee
In order to compute the change in the index as we cross this
wall, we change variables to
\be \label{edh2}
\crh_s' = (\crh_s+\cs_s+2\cv_s)/2, \quad \cs_s' = (\crh_s+\cs_s-2\cv_s)/2,
\quad \cv_s' = (\cs_s-\crh_s)/2\, .
\ee
The periodicity properties \refb{epr3} on $(\crh_s,\cs_s,\cv_s)$
take the form
\be \label{edh3}
(\crh_s',\cs_s',\cv_s') \to  (\crh_s'+2,\cs_s', \cv_s'), \,
(\crh_s',\cs_s'+2, \cv_s'), \, (\crh_s',\cs_s',\cv_s'+2), \, 
(\crh_s'+1, \cs_s'+1, \cv_s'+1)\, .
\ee
We choose the unit cell in the $(\Re(\crh_s'), \Re(\cs_s'), \Re(\cv_s'))$
to be $-1\le\Re(\crh_s')<1$, $-1\le\Re(\cs_s')<1$ and 
$-{1\over 2}\le \Re(\cv_s')
<{1\over 2}$.
Since the jacobian of the transformation associated with
\refb{edh2} is unity, the change in the index across the
wall is given by an expression analogous to 
\refb{ech1}
\be \label{edh4}
\Delta d(Q,P) =
{1\over 4}\, (-1)^{Q\cdot P+1} \, 
\int_{iM_1'-1}^{iM_1'+1}\, d\crh_s' 
\int_{iM_2'-1}^{iM_2'+1}\, d\cs_s'
\ointop d\cv_s'
 \,
e^{-i{\pi\over 4} (\cs_s' Q^{\prime 2} + \crh_s' P^{\prime 2}
+ 2 \cv_s' Q'\cdot P')} \,
{1\over \cp(\crh, \cs, \cv)}\, ,
\ee
where
\be \label{edh6}
Q' = {Q-P\over \sqrt 2}, \qquad P'={Q+P\over \sqrt 2}\, .
\ee
Substituting \refb{edh1} into \refb{edh4} we get
\ben \label{edh5}
\Delta d(Q,P) &=&
{1\over 4} 
\, (-1)^{Q\cdot P+1} \, Q'\cdot P'\, 
\int_{iM_1'-1}^{iM_1'+1}\, d\crh_s' \,
\eta(\crh_s'/2)^{-24} 
\, e^{-i\pi \crh_s' \, P^{\prime 2}/4}\, \nonumber \\
&& \qquad 
\int_{iM_2'-1}^{iM_2'+1}\, d\cs_s' \, \eta(\cs_s'/2)^{-24} 
\, e^{-i\pi \cs_s' \, Q^{\prime 2}/4} 
\, .
\een
On the other hand now the indices of the half-BPS decay
products carrying charges 
\be \label{edh5.5}
(Q_1,P_1) = ((Q-P)/2,(P-Q)/2), \qquad
(Q_2,P_2) = ((Q+P)/2,(Q+P)/2)\, ,
\ee
are given by
\be \label{edh7}
d_h(Q_1,P_1) = \int_{iM-1/2}^{iM+1/2}\, 
d\tau (\eta(\tau))^{-24}
e^{-i\pi\tau ((Q-P)/2)^2} 
= {1\over 2} \, \int_{2iM-1}^{2iM+1}\, 
d\cs_s' \, \eta(\cs_s'/2)^{-24} 
\, e^{-i\pi \cs_s' \, Q^{\prime 2}/4} 
\, ,
\ee
and
\be \label{edh8}
d_h(Q_2,P_2) = \int_{iM-1/2}^{iM+1/2}\, 
 d\tau (\eta(\tau))^{-24}
e^{-i\pi\tau ((Q+P)/2)^2} 
= {1\over 2} \, \int_{2iM-1}^{2iM+1}\, 
d\crh_s' \, \eta(\crh_s'/2)^{-24} 
\, e^{-i\pi \crh_s' \, P^{\prime 2}/4} 
\, .
\ee
Using these results and the identities
\be \label{eiden}
Q_1\cdot P_2 - Q_2\cdot P_1 = Q'\cdot P', \quad
(-1)^{Q_1\cdot P_2 - Q_2\cdot P_1}
= (-1)^{(Q-P)^2/2  -P^2 + Q\cdot P} =(-1)^{Q\cdot P}\, ,
\ee
we can express \refb{edh5} as
\be \label{edh9}
\Delta d(Q,P)
= (-1)^{Q_1\cdot P_2-Q_2\cdot P_1+1}\, 
(Q_1\cdot P_2-Q_2\cdot P_1) \,
d_h(Q_1,P_1)\, d_h(Q_2,P_2)\, ,
\ee
in agreement with the wall crossing formula.

Next consider the decay $(Q,P)\to (Q-P,0)+(P,P)$. This is controlled
by the pole at $\cs+\cv=0$. To analyze this contribution we define
\be \label{efh1}
Q' = Q-P, \quad P'=P
\ee
and
\be \label{efh2}
\crh_s'=\crh_s+\cs_s+2\cv_s, \quad \cs_s' = \cs_s, 
\quad \cv_s'=\cv_s+\cs_s\, ,
\ee
so that $\crh_sP^2+\cs_s Q^2 + 2\cv_s Q\cdot P
= \crh'_s P^{\prime 2} + \cs_s' Q^{\prime 2} + 2\cv_s'
Q'\cdot P'$.
In terms of these variables the periods are
\be \label{efh3}
(\crh_s',\cs_s',\cv_s') \to (\crh_s'+2,\cs_s',\cv_s'), \, (\crh_s',\cs_s'+1,\cv_s'), \,
(\crh_s',\cs_s',\cv_s'+2)\, .
\ee
The behaviour of $\cp$ near $\cv_s'=0$ can be found 
by the usual procedure
and result is
\ben \label{efh4}
\cp(\crh,\cs,\cv)^{-1} &=& -{1\over 4\pi^2 \cv_s^{\prime 2} }
\left\{ \eta\left({\crh_s'\over 4}\right)^{-24} 
+ \eta\left({\crh_s'+2\over 4}\right)^{-24}
\right\}\nonumber \\
&& \qquad \times \left\{ \eta\left({\cs_s'\over 4}\right)^{-24}
+ \eta\left({\cs_s'+1\over 4}\right)^{-24} + 
\eta\left({\cs_s'+2\over 4}\right)^{-24} 
+ \eta\left({\cs_s'+3\over 4}\right)^{-24}
\right\}  \nonumber \\
&& - {1\over \pi^2 \cv_s^{\prime 2}}
\, \left\{ \eta\left({\crh_s'\over 4}\right)^{-24} 
+ \eta\left({\crh_s'+2\over 4}\right)^{-24}
\right\} \, \eta(\cs_s')^{-24} +\OO\left( \cv_s^{\prime 0}\right)\, .
\een
Note that the first set of terms represent correctly the factorization
behaviour given in \refb{eabab2}, but the second set of terms
are extra. Thus the wall crossing formula gets modified for the
decay into non-primitive states.
Using \refb{efh4} we can compute 
the jump in the index across the wall
\ben \label{efh5}
&& \Delta d(Q,P)
= {1\over 16} (-1)^{Q\cdot P+1} \, Q'\cdot P' \, 
\int_{iM_1'-1}^{iM_1'+1}\, 
d\crh_s' \left\{ \eta\left({\crh_s'\over 4}\right)^{-24} 
+ \eta\left({\crh_s'+2\over 4}\right)^{-24}
\right\} e^{-i\pi\crh_s' P^{\prime 2}/4}\, 
\nonumber \\
&& 
\int_{iM_2'-1/2}^{iM_2'+1/2}\, d\cs_s' \, 
\left\{ \eta\left({\cs_s'\over 4}\right)^{-24}
+ \eta\left({\cs_s'+1\over 4}\right)^{-24} \right.\nonumber \\
&& \qquad \qquad \qquad \left. + 
\eta\left({\cs_s'+2\over 4}\right)^{-24} 
+ \eta\left({\cs_s'+3\over 4}\right)^{-24}
\right\}\,  
e^{-i\pi\cs_s' Q^{\prime 2}/4}\nonumber \\
&& + {1\over 4} (-1)^{Q\cdot P+1} \, Q'\cdot P' \, 
\int_{iM_1'-1}^{iM_1'+1}\, 
d\crh_s' \left\{ \eta\left({\crh_s'\over 4}\right)^{-24} 
+ \eta\left({\crh_s'+2\over 4}\right)^{-24}
\right\}  \, e^{-i\pi\crh_s' P^{\prime 2}/4} \nonumber \\ 
&& 
\qquad \qquad \times\, 
\int_{iM_2'-1/2}^{iM_2'+1/2}\,  d\cs_s' 
\, \eta(\cs_s')^{-24}\, e^{-i\pi\cs_s' Q^{\prime 2}/4}\, . 
\een
Defining
\be \label{efh6}
(Q_1,P_1) = (Q-P, 0), \quad (Q_2,P_2) = (P,P)\,,
\ee
we can express \refb{efh5} as
\be \label{efh7}
\Delta d(Q,P) = (-1)^{Q_1\cdot P_2 - Q_2\cdot P_1 + 1}
(Q_1\cdot P_2 - Q_2\cdot P_1)\, 
\left\{d_h\left(Q_1,P_1\right) +
d_h\left({1\over 2} Q_1,{1\over 2} P_1\right) \right\}
\, d_h\left(Q_2,P_2\right) \, .
\ee
The second term is extra compared to \refb{ex1}; it represents the
effect of non-primitivity of the final state dyons.

Finally let us turn to the analysis of the black hole entropy. For this
we need to identify the zeroes of $\cp$ at $\crh\cs-\cv^2+\cv=0$
and show that $\cp$ has the behaviour given in 
\refb{eblack6bb} near this pole. This is easily done using \refb{epr1}
and the
locations of the zeroes of $\Phi_{10}$ given in \refb{eapl5a}.
One finds that the only term that has a zero at the desired
location is the first term inside the first square bracket
in \refb{epr1}. Furthermore this term is 
proportional to the dyon partition function 
$1/\Phi_{10}(\crh,\cs,\cv)$ of the unit torsion states discussed
in \S\ref{s3.1}. Thus this
term clearly will have the desired factorization property given in
\refb{eblack6bb}.

Our proposal for the dyon partition function can be easily
generalized to the torsion 2,
primitive $Q$, $P$ and
odd $Q^2/2$,
$P^2/2$ dyons discussed at the end of \S\ref{s3.4}.  
This requires changing
the signs of appropriate terms in \refb{epr1} so that the partition
function is odd under $\crh \to \crh+{1\over 2}$ and also
under $\cs\to \cs+{1\over 2}$. The result is 
\ben \label{epr1odd}
{1\over \cp(\crh,\cs,\cv)} &=& {1\over 16} \Bigg[
{1\over \Phi_{10}(\crh,\cs,\cv)} -
{1\over \Phi_{10}(\crh,\cs+{1\over 2},\cv)} -
{1\over \Phi_{10}(\crh+{1\over 2},\cs,\cv)} \nonumber \\
&& +
{1\over \Phi_{10}(\crh+{1\over 2},\cs+{1\over 2},\cv)} +
{1\over \Phi_{10}(\crh+{1\over 4},\cs+{1\over 4},\cv+{1\over 4})} \nonumber \\
&&
-
{1\over \Phi_{10}(\crh+{1\over 4},\cs+{3\over 4},\cv+{1\over 4})} -
{1\over \Phi_{10}(\crh+{3\over 4},\cs+{1\over 4},\cv+{1\over 4})} \nonumber \\
&&+
{1\over \Phi_{10}(\crh+{3\over 4},\cs+{3\over 4},\cv+{1\over 4})} 
 +
{1\over \Phi_{10}(\crh+{1\over 2},\cs+{1\over 2},\cv+{1\over 2})} \nonumber \\
&&-
{1\over \Phi_{10}(\crh+{1\over 2},\cs,\cv+{1\over 2})} -
{1\over \Phi_{10}(\crh,\cs+{1\over 2},\cv+{1\over 2})} +
{1\over \Phi_{10}(\crh,\cs,\cv+{1\over 2})}  \nonumber \\
&&+
{1\over \Phi_{10}(\crh+{3\over 4},\cs+{3\over 4},\cv+{3\over 4})} -
{1\over \Phi_{10}(\crh+{3\over 4},\cs+{1\over 4},\cv+{3\over 4})} 
\nonumber \\
&& -
{1\over \Phi_{10}(\crh+{1\over 4},\cs+{3\over 4},\cv+{3\over 4})}  +
{1\over \Phi_{10}(\crh+{1\over 4},\cs+{1\over 4},\cv+{3\over 4})}\Bigg]
\nonumber \\
&&  +  
\Bigg[{1\over \Phi_{10}(\crh+\cs+2\cv,
\crh+\cs-2\cv,\cs-\crh)} \nonumber \\
&&
- {1\over \Phi_{10}(\crh+\cs+2\cv+{1\over 2},
\crh+\cs-2\cv+{1\over 2},\cs-\crh+{1\over 2})}
\Bigg]\, .
\een
This together with
\refb{epr1}
 exhausts all the dyons of torsion two with $Q$, $P$ primitive
since there are no dyons of this type with $Q^2/2$ even, $P^2/2$
odd or vice versa. To see this we note that since $(Q\pm P)$ are 
$2\times$
primitive vectors, $(Q\pm P)^2/2$ must be multiples of four.
Taking the sum and difference we find that
$(Q^2+P^2)/2$ and $Q\cdot P$
must be even.

Since \refb{epr1} and \refb{epr1odd} contains information about
all the torsion two dyons with primitive $(Q,P)$, the full partition
function for such dyons in obtained by taking the sum of these two
functions. This gives the result quoted in \refb{etor2}.

Given this result on torsion two dyons in string theory we can go to
appropriate gauge theory limit to extract information about torsion
two dyons in gauge theories as in \cite{0708.3715,0712.0043}. 
For simplicity we shall consider
$SU(3)$ gauge theories. If we denote by $\alpha_1$ and $\alpha_2$
the two simple roots of $SU(3)$, then, since the metric $L$ reduces
to the negative of the Cartan metric of the gauge group, we have
\be \label{egg1}
\alpha_1^2 = -2, \qquad \alpha_2^2=-2, \qquad \alpha_1\cdot
\alpha_2=1\, .
\ee
Let us now consider a dyon of charge vectors $(Q,P)$ with
\be \label{egg2}
Q=\alpha_1 - \alpha_2, \qquad P=\alpha_1+\alpha_2\, .
\ee
This has torsion 2. Furthermore both $Q$ and $P$ are primitive.
Thus this falls in the class of dyons analyzed in this section. In fact,
since
\be \label{egg3}
{Q^2 \over 2} = -3, \qquad {P^2\over 2} = -1 , \qquad 
Q\cdot P=0\, ,
\ee
the index of these quarter BPS dyons in gauge theory must be
contained in \refb{epr1odd}. 
We shall first show that the 16 terms inside the first square bracket
in \refb{epr1odd} do not
contribute to the index of the dyons described in
\refb{egg2} in any domain in the moduli space. 
For this we note that the index computed from
these terms is identical to the index of dyons of torsion 1
with appropriate constraints on $Q^2$, $P^2$ and $Q\cdot P$.
In absence of these constraints the index is 
known to reproduce the index of unit torsion gauge
theory dyons correctly\cite{0708.3715}, 
-- these are dyons of charge 
$(\alpha_1,\alpha_2)$ or ones related to it by $SL(2,\ZZZ)$ S-duality
transformation:
\be \label{egg4}
(Q,P) = (a\alpha_1+b\alpha_2, c\alpha_1+d\alpha_2), \quad
a,b,c,d\in\ZZZ, \quad ad-bc=1\, .
\ee
Such dyons will always have $Q^2P^2-(Q\cdot P)^2=3$, and
hence can never give a state of the form given in \refb{egg2} which
has $Q^2P^2-(Q\cdot P)^2=12$.
This in turn shows that the 16 terms inside the first square bracket
in \refb{epr1odd} can never give a non-vanishing contribution to the
index of a gauge theory state with charge vector given in
\refb{egg2}.

Thus the only possible contribution to the index 
of the dyons with charges 
$(\alpha_1-\alpha_2,\alpha_1+\alpha_2)$
can come from
the two terms inside the second square bracket in \refb{epr1odd}. 
In fact when
$Q^2/2$ and $P^2/2$ are odd then both 
terms give equal contribution;
so we can just calculate the contribution from the first term and
multiply it by a factor of 2. Equivalently we could use \refb{etor2}
where only the first term is present with a factor of 2.
Defining
\be \label{egg6}
\crh' = \crh+\cs+2\cv, \quad \cs'=\crh+\cs -2\cv, \quad
\cv' = \cs -\crh\, ,
\ee
we can identify the relevant term in $1/\cp(\crh,\cs,\cv)$ as
$2/\Phi_{10}(\crh',\cs',\cv')$. As usual
the contribution of this term to the index  depends on the
choice of the integration contour, which in turn is determined by
domain in the moduli space in which we want to compute the index.
Equivalently we can say that in different domains we need to
use different Fourier series expansion of
$1/\cp$. 
Now the index of a charge vector of the type given in
\refb{egg3} will come from a term in the expansion of $1/\cp$
of the form
\be \label{egg5}
e^{-2i\pi (\crh+3\cs)}\, .
\ee
Using \refb{egg6} this takes the form
\be \label{egg7}
e^{-2i\pi(\crh'+\cs'+\cv')}\, .
\ee
Thus in whichever domain the Fourier
expansion of $1/\cp$ contains a term of the form \refb{egg7} we
have a non-vanishing index for the dyon in \refb{egg3} with the
index being equal to $(-1)^{Q\cdot P+1}$ times
the coefficient of this term. 
Since here $Q\cdot P=0$, the index is $-2$ times the coefficient
of \refb{egg7} in the Fourier
expansion of $1/\Phi_{10}(\crh',\cs',\cv')$.
Now from the analysis of partition function of torsion one dyons
(see {\it e.g.} \cite{0708.3715}) 
we know that this expansion indeed
has  a term of the form \refb{egg7} for one class of choices
of contour; these are the contours for which 
\be \label{egg8}
\Im(\crh'), \Im(\cs') >> -\Im(\cv') >> 0 \, .
\ee
Using \refb{echoice}, or equivalently by an $SL(2,\RRR)$
transformation of the results of \cite{0708.3715}, one can 
figure out the domain in the moduli space in
which this choice of contour is the correct one. 
It turns out to be the domain bounded by the walls associated with the
decays of $(Q,P)$ into 
\be \label{ewallgauge}
((Q-P)/2,(P-Q)/2) + ((Q+P)/2,(Q+P)/2), \quad (-P,P)+(P+Q,0), \quad
(Q-P,0)+(P,P)\, .
\ee
In this domain
$1/\Phi_{10}$ has to be first expanded in powers of $e^{2\pi i\crh'}$
and $e^{2\pi i\cs'}$ and then each coefficient needs to be expanded
in powers of $e^{-2\pi i\cv'}$.
\refb{egg7} is the  leading term in this
expansion  and its coefficient
is 1. As a result the
index of the dyons is $-2$. This agrees 
with the results of \cite{9804174,9907090,0005275,0609055}
where it was shown that in an appropriate domain in the
moduli space dyons of torsion $r$ has index $(-1)^{r-1} r$,
since these dyonic states are obtained by tensoring the basic
supermultiplet with a state of spin $(r-1)/2$.

Using a string junction picture\cite{9712211,9804160}
ref.\cite{9804174} also showed that the dyon
considered above exists in a domain of the moduli space bounded
by three walls of marginal stability, -- one asscociated with the decay
into $(\alpha_1,\alpha_1)+(-\alpha_2,\alpha_2)$, the second
associated with the decay into $(2\alpha_1,0)
+(-\alpha_1-\alpha_2, \alpha_1+\alpha_2)$ 
and the third one associated
with the decay into 
$(-2\alpha_2,0)+(\alpha_1+\alpha_2,\alpha_1+\alpha_2)$.
These are precisely the walls listed in \refb{ewallgauge}. Since
the gauge theory dyons cease to exist outside these walls, 
the index computed in gauge theory jumps by $2$ across
these three walls of marginal stability. 
Does this agree with the prediction of the proposed dyon
partition function in string theory?
We can
calculate the change in the index associated with these
decays using the standard formula \refb{ex1} for decay into primitive
dyons\cite{appear} and the 
modified formula \refb{efh7} for decay into
non-primitive dyons since the proposed partition function satisfies
these relations. We find a jump in the index equal to 2 across each of these
walls
as predicted by the gauge theory results.

\sectiono{Reverse applications} \label{sapp}

In our analysis so far we have used the information on half BPS
partition function to extract information about quarter BPS
partition function. However we can turn this around. If the
quarter BPS partition function is known then we can use it to
extract information about the half-BPS partition function by
first identifying  an appropriate wall on which one of the decay
products is the half BPS state under consideration and then studying
the behaviour of the quarter BPS partition function near the
pole that controls the jump in the index at this particular wall.

As an example we can consider the $\ZZZ_2$ CHL model of
\S\ref{s3.3}. The decay $(Q,P)\to (Q,0)+(0,P)$ is controlled
by the behaviour of $\cp$ near $\cv=0$. Thus if we did not know
the spectrum of magnetically charged half BPS states in this theory,
we could study the behaviour of $\cp$ near $\cv=0$ to get this
information. In this case since all the other walls are related
to this one by S-duality transformation, this is the only
independent information we can get.  However for more
complicated models there can be more information.

To illustrate this we shall consider the example of the $\ZZZ_6$
CHL model\cite{9508144,9508154} 
mentioned in footnote \ref{ff1}. Our set $\AAA$
consists of charge vectors of the form
\be \label{eapl1}
Q=\pmatrix{0 \cr m/6\cr 0 \cr -1}\, , \qquad P = 
\pmatrix{K \cr J \cr 1\cr 0}\, , \quad m, K, J\in\ZZZ\, ,
\ee
as in \refb{ed1}. We now consider the decay associated with the
matrix
\be \label{eapl2}
\pmatrix{a_0 & b_0\cr c_0 & d_0} = \pmatrix{1 & 1\cr 2 & 3}\, .
\ee
{}From \refb{ede1} we see that this corresponds to the decay
\be \label{eapl3}
(Q,P) \to (M_0, 2M_0) + (N_0, 3N_0) \, , \quad
M_0\equiv 3Q-P, \quad N_0\equiv -2Q + P\, .
\ee
The charge vectors $M_0$ and $N_0$ are not related to $Q$ or $P$
by a T-duality transformation since they correspond to charges
that are triple and double twisted respectively. Furthermore
the dyon charges $(M_0,2M_0)$ and $(N_0,3N_0)$ cannot be
related by S-duality group $\Gamma_1(6)$ to either a purely
electric or a purely magnetic state whose index is known.
On the other hand the partition function of quarter BPS states of
the type given in \refb{eapl3} is known\cite{0609109,0708.1270}.
Thus the latter can be used to extract information about the
partition function of these half BPS states.

{}From \refb{epole1} it follows that the relevant zero of
$\cp$ we need to examine is at
\be \label{eapl4}
6\crh + \cs + 5\cv = 0\, .
\ee
The zeroes of $\cp$ have been classified in 
\cite{0605210,0708.1270}.
For a generic $\ZZZ_N$ model $\cp$ has double zeroes at
\ben \label{eapl5}
&& n_2 ( \cs \crh  -\cv ^2) + j\cv  + 
n_1 \cs  -\crh m_1 + m_2 = 0 \nonumber \\
&& m_1\in N\ZZZ, \quad
n_1, m_2, n_2 \in \ZZZ, \quad j\in 2\ZZZ+1, \quad
m_1 n_1 + m_2 n_2 +\frac{j^2}{4} = {1\over 4}\, .
\een
For the $N=6$ model, taking 
\be \label{eapl6}
m_1 = -6, \quad n_1=1, \quad m_2 = n_2=0, \quad j = 5\, ,
\ee
we see that $\cp$ indeed has a zero at \refb{eapl4}. Thus by
examining the known expression for $\cp$ near this zero
we can determine the half-BPS partition functions of interest.
This can be done in a straightforward manner following the
general procedure described in \cite{0605210,0708.1270}.

We note in passing that
in an arbitrary $\ZZZ_N$ model with the set $\AAA$ chosen as
\be \label{eapl7}
Q=\pmatrix{0 \cr m/N\cr 0 \cr -1}\, , \qquad P = 
\pmatrix{K \cr J \cr 1\cr 0}\, , \quad m, K, J\in\ZZZ\, ,
\ee
the walls of marginal stability are controlled by matrices
$\pmatrix{a_0 & b_0\cr c_0 & d_0}$ subject to the 
conditions\cite{0702141}
\be \label{eapl8}
a_0d_0-b_0 c_0=1, \quad 
a_0, b_0, c_0, d_0\in\ZZZ, \quad 
c_0 d_0\in N\ZZZ\, .
\ee
According to our hypothesis this decay will be controlled by a
double zero of $\cp$ at
\be \label{eapl9}
\crh c_0d_0 + \cs a_0b_0 + \cv (a_0d_0 + b_0c_0) = 0\, .
\ee 
This corresponds to the choice
\be \label{eapl10}
m_1 = - c_0 d_0, \quad n_1 = a_0 b_0 , \quad m_2 = n_2 =0, \quad
j = a_0d_0 + b_0c_0\, ,
\ee
in eq.\refb{eapl5}.
We now see that the $m_i$'s, $n_i$'s and $j$ given in \refb{eapl10}
satisfies all the constraints mentioned in \refb{eapl5} as a 
consequence of \refb{eapl8}. Thus our proposal that the
decay associated with the matrix $\pmatrix{a_0 & b_0\cr
c_0 & d_0}$ is always controlled by the zero at 
\refb{eapl9} is at least consistent with the locations of the zeroes
of $\cp$ for $\ZZZ_N$ orbifold models.

\medskip

\noindent {\bf Acknowledgment:} We would like to thank
Nabamita Banerjee, Miranda Cheng, Atish Dabholkar,
Justin David, 
Frederik Denef, Dileep Jatkar, Suresh Nampuri and
K. Narayan for useful discussions.


\end{document}